\documentclass[fleqn,usenatbib]{mnras}

\usepackage{newtxtext,newtxmath}

\usepackage[T1]{fontenc}
\usepackage{ae,aecompl}

\usepackage{graphicx}	
\usepackage{amsmath}	
\usepackage{xcolor}
\usepackage{hyperref}
\usepackage[normalem]{ulem}     
\usepackage{multirow} 
\usepackage{siunitx} 

\title[\namecode{} map-making code for the QUIJOTE experiment.]{The \namecode{} map-making code: application to a simulation of the QUIJOTE northern sky survey}

\author[Guidi et al.]{F. Guidi,$^{1,2}$\thanks{E-mail: fguidi@iac.es} J.~A. Rubi{\~n}o-Mart{\'{\i}}n,$^{1,2}$\thanks{E-mail: jalberto@iac.es} A.~E. Pelaez-Santos,$^{1,2}$ R.~T. G{\'e}nova-Santos,$^{1,2}$ \newauthor M. Ashdown,$^{6,7}$ R.~B. Barreiro,$^{3}$ J.~D. Bilbao-Ahedo,$^{3,4}$  S.~E. {Harper},$^{5}$ R.~A. {Watson}$^{5}$  \\
\\
$^{1}$Instituto de Astrof\'{\i}sica de Canarias, E-38200 La Laguna, Tenerife, Spain\\
$^{2}$Departamento de Astrof\'{\i}sica, Universidad de La Laguna,
E-38206 La Laguna, Tenerife, Spain\\
$^{3}$Instituto de F\'{\i}sica de Cantabria (IFCA), CSIC-Univ. de Cantabria, Avda. los
Castros, s/n, E-39005 Santander, Spain\\
$^{4}$Departamento de F\'{\i}sica Moderna, Universidad de Cantabria, Avda. de los Castros s/n, E-39005 Santander, Spain \\
$^{5}$Jodrell Bank Centre for Astrophysics, Alan Turing Building, University of Manchester, Manchester M13 9PL, UK\\
$^{6}$Astrophysics Group, Cavendish Laboratory, University of Cambridge, J J Thomson Avenue, Cambridge CB3 0HE, UK\\
$^{7}$Kavli Institute for Cosmology, University of Cambridge,
Madingley Road, Cambridge CB3 0HA, UK\\
}

\date{Accepted XXX. Received YYY; in original form ZZZ}

\pubyear{2021}

\begin{document}



\newcommand\oof{$1/f$ }
\newcommand\mk{\,mK}
\newcommand\QJT{{\it QUIJOTE}}
\newcommand\jkn{Jack-Knife}
\newcommand\cl{$C_{\ell}$}
\newcommand\namecode{{\sc PICASSO}}
\newcommand\xpol{{\sc Xpol}}

\label{firstpage}
\pagerange{\pageref{firstpage}--\pageref{lastpage}}
\maketitle

\begin{abstract}
Map-making is an important step for the data analysis of Cosmic Microwave Background (CMB) experiments. It consists of converting the data, which are typically a long, complex and noisy collection of measurements, into a map, which is an image of the observed sky.
We present in this paper a new map-making code named \namecode{} (Polarization and Intensity CArtographer for Scanned Sky Observations), which was implemented to construct intensity and polarization maps from the Multi Frequency Instrument (MFI) of the QUIJOTE (Q-U-I Joint TEnerife) CMB polarization experiment.  \namecode{} is based on the destriping algorithm, and is suited to address specific issues of ground-based microwave observations, with a technique that allows the fit of a template function in the time domain, during the map-making step. 
This paper describes the \namecode{} code, validating it with simulations and assessing its performance. For this purpose, we produced realistic simulations of the QUIJOTE-MFI survey of the northern sky (approximately $\sim 20,000$\,deg$^2$), and analysed the reconstructed maps with \namecode{}, using real and harmonic space statistics. 
We show that, for this sky area, \namecode{} is able to reconstruct, with high fidelity, the injected signal, recovering all the scales with $\ell>10$ in TT, EE and BB. The signal error is better than 0.001\,\% at $20<\ell<200$.
Finally, we validated some of the methods that will be applied to the real wide-survey data, like the detection of the CMB anisotropies via cross-correlation analyses. 
Despite that the implementation of \namecode{} is specific for QUIJOTE-MFI data, it could be adapted to other experiments.
\end{abstract}

\begin{keywords}
methods: data analysis - cosmology: observations -  cosmic background radiation - diffuse radiation
\end{keywords}



\section{Introduction} 
The first product of scanned observations at radio and microwave frequencies is not an image of the observed sky, but a Time Ordered Data stream (TOD) \citep[e.g.][]{Tegmark1997}. The TOD is a long collection of sky signal measurements, recorded by the instrument as a function of time and pointing coordinates, in combination with a component of instrumental and atmospheric noise. The role of the map-making is to project and integrate this set of measurements from the time domain to their original sky positions, and to construct a map, which is an image of the observed sky. The map is a compressed version of the data, and can be analyzed in the harmonic space by computing its angular power spectrum, which is also one of the main observables for CMB experiments.


This work has been developed in the context of the QUIJOTE\footnote{Web page: \url{http://research.iac.es/project/cmb/quijote}} CMB experiment \citep{RubinoSPIE12}, a ground-based polarimeter installed at the Teide Observatory\footnote{\url{http://www.iac.es/es/observatorios-de-canarias/observatorio-del-teide}} (Tenerife, Spain), with the aim of studying the linearly polarized microwave sky at $\approx 1^{\circ}$ angular resolution. The MFI instrument of QUIJOTE \citep{MFIstatus12} is sensitive to the frequency range $10$--$20$\,GHz, with four central frequencies at $11,\,13,\,17,\,19$\,GHz. The first results based on QUIJOTE-MFI instrument can be found in \cite{Perseus, W44,Taurus}. The MFI concluded in 2018 its observational campaign, during which it also carried out a survey of the northern sky: the QUIJOTE-MFI wide-survey (Rubi{\~n}o-Mart{\'{\i}}n et al., in prep.). 

The wide-survey maps obtained with QUIJOTE-MFI data, in combination with other low frequency surveys like C-BASS (\citealt{Jones2018}) and S-PASS (\citealt{SPASS}), can be used to complement the Planck (\citealt{Planck2018}) and WMAP (\citealt{WMAPmaps}) missions, to improve component separation for any present and future CMB experiment. The QUIJOTE-MFI survey, in particular, provides a precise characterization of the low frequency Galactic foregrounds, like the anomalous microwave emission (AME) and the synchrotron. The scientific analysis of the wide-survey maps will be presented in a set of papers in preparation (Poidevin et al.; Vansyngel et al.; Ruiz-Granados et al.; Watson et al.; de la Hoz et al.; Herranz et al.; Tramonte et al.; Guidi et al.; Fern\'andez Torreiro et al., L\'opez-Caraballo et al.) and in the already published \citealp{Lambda_Orionis}. 

In this paper, we present the \namecode{} map-making code, which was implemented for the construction of the QUIJOTE-MFI intensity and polarization maps. In particular, \namecode{} has been already used to produce maps from raster scans observations in some Galactic regions \citep{W44,Taurus}, and will be applied to the aforementioned wide-survey maps. 
The \namecode{} code is based on the destriping algorithm implemented in the {\sc MADAM} code, which was used for the  construction of the Planck-LFI maps (\citealt{Madam2}, \citealt{Madam},  \citealt{Planck-mapmaking}), and which is commonly used for the map-making of CMB experiments  (\citealt{Multidetector}, \citealt{Kurki-Suonio}, \citealt{Planck_LFI_30}).
Our code  is an independent implementation, and is suited to address specific issues related with QUIJOTE, or, in general, with other ground-based microwave experiments. 

Furthermore, \namecode{} has been implemented with a technique that is useful for ground-based experiments, with the aim to fit and subtract from the data a general template at the TOD level, during the map-making step. We applied this technique for the subtraction of two particular templates which are important at microwave wavelengths: the CMB dipole, and a stable, plane-parallel atmospheric component, but it can also be particularly useful for ground-based experiments, for example for the correction of radio frequency interference (RFI). 

Here we validate \namecode{}, showing its performance with realistic end-to-end simulation of the QUIJOTE-MFI wide-survey data. We employed TOD simulations containing foreground sky signal, point sources, CMB anisotropies, CMB dipole, and three different scenarios for the noise: no-noise, only white noise, and correlated noise (including realistic \oof plus a white noise component). We reconstructed maps of the simulated data-set with \namecode{}, and we studied their angular power spectra, in order to quantify the ability of the code to reconstruct the injected sky signal. Particular attention was placed on the characterization of the large angular scales signal. 

In addition, the simulations provide a useful tool for the validation of some of the results obtained with the real wide-survey maps. We discuss in this work the detection of the CMB anisotropies in intensity through cross-correlations using the simulation, as a support for the methodology and the result obtained with the real data, that will be presented in Rubi{\~n}o-Mart{\'{\i}}n et al. (in prep.).

The paper is organized as follows: in Section~\ref{sec_destriper} we describe the generalities of the destriper map-making algorithm, in Section~\ref{sec_QJT_mapmaking} we present the specifics of \namecode{} suited for QUIJOTE-MFI data, and in Section \,\ref{sec_fitfunc} the implementation of template function fitting at the map-making level. Then, in Section~\ref{sec_code} we briefly describe the structure of the code, and in Section~\ref{Sec_simu} we present the realistic end-to-end simulations of the QUIJOTE-MFI wide-survey, that where used to validate the map-making procedure. Finally, the results are reported in Section~\ref{sec_results}, where we show the maps of the simulations, the analysis at the power spectrum level, the transfer function, the cross-correlations with the CMB anisotropies, and two examples of the fitting of a template function, using a static atmosphere and the CMB dipole. We report our conclusions in Section~\ref{sec_conclusions}.

\section{Map-making problem} 
\label{sec_destriper}

The map-making problem consists of finding an efficient and optimal way to project the TOD into a map of the observed sky, by accounting simultaneously for the suppression of the correlated noise. Different techniques have been presented in the literature, and they are mainly based on the maximum-likelihood  (e.g., \citealp{Tegmark1997}) and destriping (e.g.,  \citealp{Delabrouille1998,Burigana1999,Maino1999,Madam2}) techniques. In addition, different filtering operations at the map-making level have been proposed (e.g., \citealp{Poletti2017}), with the aim of reducing unwanted noise modes from the data.

\namecode{} is based in the destriping technique, which is widely used in the context of CMB experiments (\citealt{Kurki-Suonio, Madam, Multidetector, Planck-mapmaking}). All microwave experiments, indeed, have the common goal of obtaining the cleanest possible intensity and polarization maps from data that are affected by correlated \oof noise, but without suppressing the large angular scale modes coming from sky signal. Usually, the TOD of a CMB experiment is  contaminated by two noise components: the white noise, which is uncorrelated, and the \oof noise that is correlated in time.  The white noise is produced by random thermal fluctuations of the electrons in the low noise amplifiers, it is Gaussianly distributed with zero mean and variance $\sigma^2$. On the other hand, the \oof noise is correlated in time, and it consists of long-time drifts which are mainly produced by instrumental gain variation and atmospheric emission.
The destriping technique aims to correct for the correlated noise component by modeling the \oof drifts with a set of consecutive offsets with a determined time length $t_{\rm b}$, the so called baselines. 

Our reference implementation of the destriping problem is the MADAM code (\citealp{Madam2, Madam}), which is implemented with priors on the baselines, taking the advantage of the a priori statistical knowledge of the noise of the experiment. Here we describe the mathematics that is at the basis of the destriping algorithm, which is then adapted and expanded for the production of the maps of the QUIJOTE-MFI experiment, in the \namecode{} map-making code.

\subsection{Destriper algorithm} 
\label{sec:destr_algorithm}

The map-making problem requires a solution for the sky map, $\textbf m_{\rm sky}$, given the detector TOD, $\textbf{y}$, which contains $n_{\rm t}$ data samples.
For a experiment measuring intensity and linear polarization like QUIJOTE, the sky map $\textbf m_{\rm sky}$ is represented as a set of three HEALPix\footnote{ \url{https://sourceforge.net/projects/healpix/} \label{footnote_healpix}} (\citealt{Gorski2005}) vectors of $n_{\rm p}$ pixels: the \textit{I}, \textit{Q} and \textit{U} Stokes maps. 
In general, the TOD vector can be written as a combination of sky signal $\textbf{s}$, and noise $\textbf n$, as:
\begin{equation}
    \textbf{y}=\textbf{s}+\textbf{n}=\textbf{P} \cdot \textbf m_{\rm sky}+\textbf{n}
    \label{y=s+n}
\end{equation}
being $\textbf{s}$ the sky signal map $\textbf m_{\rm sky}$ projected into the time ordered domain by the pointing matrix $\textbf{P}$ (see Sec.~\ref{sec_proj_mat} for the definition of QUIJOTE-MFI pointing matrix). 

The destriping algorithm describes the noise vector $\textbf{n}$ in the TOD as the sum of two components: a white noise (uncorrelated) part $\textbf{w}$, plus a correlated component, usually ascribed to the \oof noise, modelled as a series of some base functions. For this correlated part, it is commonly used a set of discrete offsets, called baselines. Thus, the noise vector is written as
\begin{equation}
  \textbf{n}=\textbf{w}+\textbf{F}\cdot \textbf{a},
  \label{n=fa}
\end{equation}
where the term $\textbf{F}\cdot \textbf{a}$ is an approximation to model the correlated \oof noise as a sequence of $n_{\rm b}$ baselines $\textbf a$, which are projected in a TOD format with the baselines pointing matrix $\textbf F$. 
We call $\textbf{C}_{\rm n}$, $\textbf{C}_{\rm w}$, and $\textbf{C}_\text{a}$ the covariance matrices of the total, white, and $1/f$ noise components, respectively (see Sec.~\ref{sec_cov} for extended description). Combining Eq.~\ref{y=s+n} and \ref{n=fa}, the data vector can be re-written as:
\begin{equation}
    \textbf{y}=\textbf{P} \cdot \textbf{m}_{\rm sky}+\textbf{F}\cdot \textbf{a}+\textbf{w}
    \label{y=fa+w+pm}
\end{equation}
and can be treated with a Bayesian statistical approach, where the parameters are the baselines vector $\textbf{a}$ and sky map $\textbf{m}_{\rm sky}$.

\noindent
The posterior of the map-making problem is given by:
\begin{equation}
P(\textbf m_{\rm sky},\textbf a|\textbf{y})  \varpropto P(\textbf m_{\rm sky},\textbf a)\cdot L(\textbf{y}),
\label{eq:likelihood0}
\end{equation}
where $P(\textbf m_{\rm sky},\textbf a)$ is the prior of the parameters, and $L(\textbf{y})=P(\textbf{y}|\textbf m_{\rm sky},\textbf a)$ is the likelihood function. If we apply
the probability product rule to Eq.~\ref{eq:likelihood0}, we get:
\begin{align}
\notag
P(\textbf m_{\rm sky},\textbf a|\textbf{y})\propto &
P(\textbf a|\textbf m_{\rm sky})P(\textbf m_{\rm sky})\cdot L(\textbf{y})= \\ 
= & P(\textbf a)P(\textbf m_{\rm sky})\cdot L(\textbf{y})
\label{posterior}
\end{align}
where $P(\textbf m_{\rm sky})$ is the prior on the map, and $P(\textbf a|\textbf m_{\rm sky})=P(\textbf a)$ is the prior
on the baselines, where we assume that the baselines values are independent from the sky signal.  To avoid imposing any prior on the map, we use a flat prior for $\textbf m_{\rm sky}$:
\begin{equation}
    P(\textbf m_{\rm sky}) = 1 \label{p_sky}
\end{equation} 
For the baselines, instead,  we assign a Gaussian prior given by:
\begin{equation}
P(\textbf a)=\frac{1}{((2\pi)^{n_{\rm b}} \det(\textbf{C}_{\text a}))^{1/2}}\exp\left(-\frac{1}{2}\textbf a^{T}\cdot \textbf{C}_{\text a}^{-1}\cdot \textbf a\right)\label{eq:Prior_base}
\end{equation}
where we assume that the baselines have a random and Gaussian (zero centered) distribution. 

\noindent
The likelihood of the data $\textbf{y}$, using Eq.~\ref{y=fa+w+pm}, is given by:
\begin{equation}
L(\textbf{y})=P(\textbf{y}|\textbf m_{\rm sky},\textbf a)=\frac{1}{((2\pi)^{n_{\rm t}} \det(\textbf{C}_{\rm w}))^{1/2}}\exp\left(-\frac{1}{2}\textbf{w}^{T}\cdot \textbf{C}_{\rm w}^{-1}\cdot\textbf{w}\right)\label{eq:like_data}
\end{equation}
where $\textbf{C}_{\rm w}$ is the covariance matrix of the white noise defined in Eq.~\ref{cov_white}.
Let us now maximize the posterior in  Eq.~\ref{posterior}, or equivalently, we minimize the negative of its logarithm $X$:\begin{equation}
X =  -2\ln\left[P(\textbf m_{\rm sky},\textbf a|\textbf{y})\right]
\end{equation}
that, using Eq.~\ref{y=fa+w+pm} and the probability density functions in Eq.~\ref{p_sky},~\ref{eq:Prior_base} and ~\ref{eq:like_data}, can be written as:
\begin{equation}
     X=(\textbf{y}-\textbf P \cdot \textbf m_{\rm sky}-\textbf F \cdot \textbf a)^{T}\textbf{C}_{\rm w}^{-1}(\textbf{y}-\textbf P  \cdot \textbf m_{\rm sky}-\textbf F  \cdot \textbf a)+\textbf a^{T}\textbf{C}_{\text a}^{-1}\textbf a+{\rm const}
      \label{eq:X}
\end{equation}
The minimization of $X$  with respect to $\textbf m_{\rm out}$ provides the destriper solution for the map:
\begin{equation}
\textbf m_{\rm out}=\textbf{M}^{-1}\textbf{P}^{T}\textbf{C}_{\rm w}^{-1}(\textbf{y}-\textbf F\cdot \textbf a) \label{mout}
\end{equation}
being $\textbf{M}=\textbf{P}^{T}\textbf{C}_{\rm w}^{-1}\textbf{P}$. We can use now Eq.~\ref{mout} to maximize $X(\textbf m=\textbf m_{\rm out})$ with respect to $\textbf{a}$, and obtain the equation to determine the baselines $\textbf a_{\rm out}$, which is:
\begin{equation}
    \left({\bf D}+\textbf{C}_{\text a}^{-1}\right) \cdot \textbf a_{\rm out} = \textbf{F}^{T}\textbf{C}_{\rm w}^{-1}\textbf{Z}\textbf{y} \label{aout}
\end{equation}
where we defined the $(n_{\rm t},n_{\rm t})$ matrix: 
\begin{equation}
    \textbf{Z}=\textbf{1}-\textbf{P}(\textbf{M}^{-1}\textbf{P}^{T}\textbf{C}_{\rm w}^{-1})
\end{equation}
and the $(n_{\rm b},n_{\rm b})$ matrix:
\begin{equation}
    \textbf{D}=\textbf{F}^{T}\textbf{C}_{\rm w}^{-1}\textbf{Z}\textbf{F}.
\end{equation}
Equations~\ref{mout} and~\ref{aout} constitute the solution the map-making problem: with Eq.~\ref{aout} we can estimate the baselines, and with Eq.~\ref{mout} we project the destriped data $(\textbf y - \textbf F \cdot \textbf a_{\rm out})$ into the the \textit{I}, \textit{Q}, and \textit{U} maps.

\subsection{Noise covariance matrices}\label{sec_cov} 

We define here the covariance matrix of the noise, distinguishing two different components: the white noise and the correlated \oof noise, modelled with the baselines.
The $(n_{\rm t} \times n_{\rm t})$ covariance matrix of the total noise is defined as:
\begin{equation}
     \textbf{C}_{\rm n}=\left<\textbf n\cdot \textbf n^{T}\right>
     \label{cn}
\end{equation}
where $<>$ indicates the ensemble average. $\textbf{C}_{\rm n}$ can be expressed as a combination of the covariance of the white and \oof noise components:
\begin{equation}
    \textbf{C}_{\rm n}=\textbf F \textbf{C}_{\text a} \textbf F^T+\textbf{C}_{\rm w}
\end{equation}
In Fourier space, the covariance of the white plus \oof{}noise is expressed in terms of the power spectral density as a function of the frequency $f$. It is usually written as (\citealp{Planck2018}):
  \begin{equation}
    P(f)= \frac{\sigma^{2}}{f_{\rm s}}\left(1+ \left(\frac{f_{\rm k}}{f}\right)^\gamma\right)
    \label{ps_teo_whiteplusoof}
  \end{equation}
where $\sigma$ sets the white noise level (see Eq.\,\ref{cov_white}), $f_{\rm s}$ is the sampling frequency of the data, $f_{\rm k}$  and $\gamma$ are, respectively, the knee frequency and the power of the \oof noise slope. 

The covariance matrix of the white noise is the $(n_{\rm t} \times n_{\rm t})$ diagonal matrix:
\begin{equation}
  C_{{\rm w},ij}=\left<w_{i}\cdot w_{j}^{T}\right>=\delta_{ij}\sigma^{2}
  \label{cov_white}
\end{equation} with $i,j=1,...,n_{\rm t}$, and where $\sigma$ is the standard deviation of the white noise component. 

\begin{figure}
  \centering
  \includegraphics[width=0.5\textwidth]{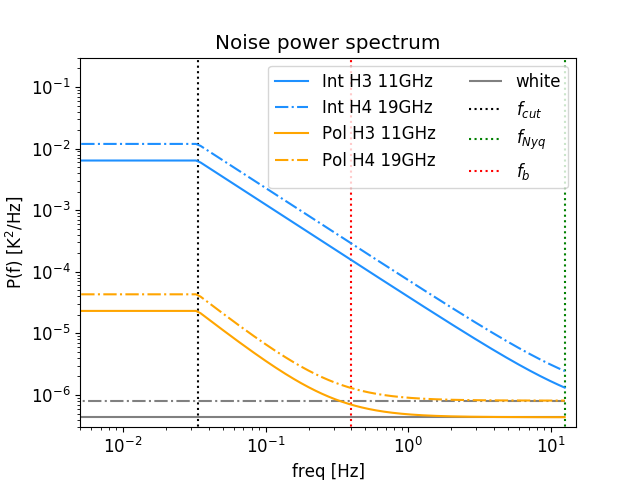}
  \caption{White plus $1/f$ power spectral density of the QUIJOTE-MFI simulated (injected) noise, for intensity (in blue) and polarization (in orange). The noise model is given by Eq.\,\ref{ps_teo_whiteplusoof}, computed for the typical values that are listed in Table~\ref{Table_noise_sim}, and where we applied a low frequency cut at $f_{\rm cut}=1/30\,{\rm s}^{-1}$ as in Eq.\,\ref{ps_teo} (including the white noise contribution).  We show the cases of two representative channels of the MFI: horn number 3 at 11\,GHz (thick lines), and horn number 4 at 19\,GHz (dot-dashed lines), and the corresponding white noise level (grey lines). We represent with vertical dotted lines the typical frequency thresholds: the low frequency threshold $f_{\rm cut}$ in black, the frequency of the baselines $f_{\rm b}=1/t_{\rm b}$ in red, with $t_{\rm b}=2.5$\,s (see Sec.~\ref{sec:destriping}), and the Nyquist frequency of the data $f_{\text{Nyq}}=1/(2\cdot t_{\rm s})$ in green, where $t_{\rm s}=40$\,ms is the sampling interval of the data (see Sec.~\ref{sec_instru_model}). Note that the knee frequency for intensity exceeds the Nyquist frequency imposed by the sampling rate.} \label{img_ps}
\end{figure}

\begin{figure}
  \centering
  \includegraphics[width=0.5\textwidth]{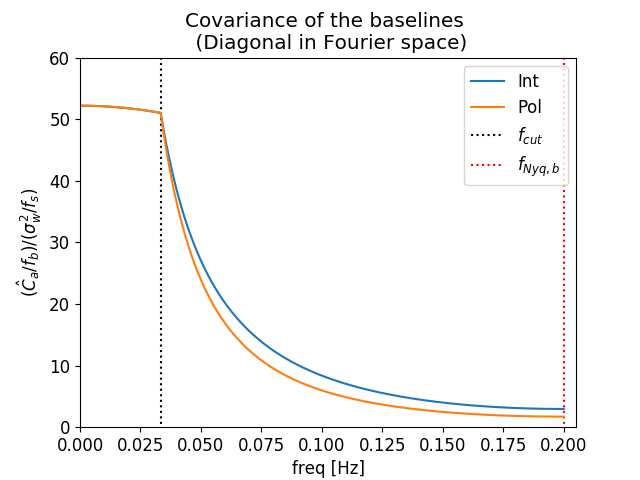}
  \caption{Diagonal of the covariance matrix of the baselines in Fourier space (Eq.\,\ref{Ca_full}) as function of frequency, for the QUIJOTE \oof priors parameters in intensity (blue) and polarization (orange), normalized to a baseline frequency of $f_{\rm b}=1$\,Hz, and to a white noise level of 1 (see Table~\ref{Table_noise_sim}). For display purposes, the amplitude of $\hat{C}_{\text{a}}(\mathrm{freq}=0)$ in intensity is normalized to the level of the polarization. The black dotted line shows the low frequency threshold $f_{\rm cut}$ of the power spectrum of the noise (see Eq.\,\ref{ps_teo} and Fig.\,\ref{img_ps}), and the red dotted line marks the Nyquist frequency of the baselines $f_{\text{Nyq},\,{\rm b}}=1/(2\cdot t_{\rm b})$, with $t_{\rm b}=2.5$\,s.} \label{img_aps}
\end{figure}

The covariance matrix of the \oof noise can be given in terms of the covariance matrix of the baselines $\textbf{C}_{\text a}$, which is a projection in the $(n_{\rm b} \times n_{\rm b})$ space of the total noise covariance $\textbf{C}_{\rm n}$ (\citealt{Multidetector}):
\begin{equation}
\textbf{C}_{\text a}  =  \left<\textbf a^{T}\textbf a\right>=\left(\textbf{F}^{T}\textbf{F}\right)^{-1}\textbf{F}^{T}\textbf{C}_{\rm n}\textbf{F}\left(\textbf{F}^{T}\textbf{F}\right)^{-1}=\frac{1}{n_{\rm b'}^{2}}\textbf{F}^{T}\textbf{C}_{\rm n}\textbf{F} \label{Ca_def0}
\end{equation}
with $n_{\rm b'}$ the number of data samples in one baseline.
Due to the correlation of the noise, $\textbf{C}_{\text a}$ is not diagonal in the time domain. However, within a good approximation, $\textbf{C}_{\text a}$ is a circulant matrix, and the correspondent matrix in Fourier space, $\hat{\bf C}_{\text{a}}$, is diagonal. 
The diagonal of $\hat{\bf C}_{\text{a}}$ can be estimated from the power spectral density of the \oof correlated noise, that can be written, similarly to Eq.\,\ref{ps_teo_whiteplusoof}, as:
  \begin{equation}
    P(f)= 
    \begin{cases}
       \frac{\sigma^{2}}{f_{\rm s}}\big( \frac{f_{\rm k}}{f}\big)^\gamma, & \text{if}\,\, f>f_{\rm cut} \\
      
   \frac{\sigma^{2}}{f_{\rm s}}\big( \frac{f_{\rm k}}{f_{\rm cut}}\big)^\gamma, & \text{if} \,\,f<f_{\rm cut} 
    \end{cases}
    \label{ps_teo}
  \end{equation}
where $f_{\rm cut}$ is a parameter that signs the transition between the \oof and a flat regime at low frequencies (e.g., in Fig.~\ref{img_ps}). 
The diagonal of the covariance matrix of the
baselines in the Fourier space is given by (\citealt{Madam}):
\begin{equation}
\hat{C}_{\text{a},ii}=P_{\rm a}(f_{i})=\frac{1}{t_{\rm b}}\sum_{m=-\infty}^{+\infty}P\left(f_{i}+\frac{m}{t_{\rm b}}\right)\frac{\sin^{2}(\pi(f_{i}t_{\rm b}+m))}{(\pi(f_{i}t_{\rm b}+m))^{2}}\label{Ca_full}
\end{equation}
where $P_{\rm a}$ is the spectrum of the baselines, $P(f_j)$ is the power spectral density of the correlated noise (Eq.\,\ref{ps_teo}) estimated at the  discrete frequencies $f_j=j/(n_{\rm b} \cdot t_{\rm b})$, with $j=(0,...,n_{\rm b}/2)$, where we remind that $t_{\rm b}$ is the time length of the baselines,  $n_{\rm b}$ is the number of baselines in the TOD, and that the highest frequency in which we compute the power spectral density is the Nyquist frequency in the baselines space. We represent an example of the diagonal of $\hat{\bf C}_{\text{a}}$ in Fig.~\ref{img_aps}.

\begin{table}  
    \centering
    \begin{tabular}{cccccc}
      \hline \hline
        & $\sqrt{\sigma^2/f_{\rm s}}$ &  $f_{\rm k}$ &  $f_{\rm k}$ (prior) & $\gamma$ &  $f_{\rm cut}$ \\  
        & $[{\rm mK} \cdot {\rm s}^{-1/2} ]$ &  [Hz] &  [Hz]  &  &  [Hz] \\ \hline \hline
          I &    0.66-0.90(11-19\,GHz)    & 20.0   & 40.0     & 1.5      & 0.033       \\ \hline
         QU &     0.66-0.90(11-19\,GHz)  & 0.3   &0.3      & 1.8      & 0.033        \\ \hline\hline
      \end{tabular}
    \caption{Reference values for the \oof{}noise parameters adopted in this paper, for a noise power spectral density of the type of Eq.~\ref{ps_teo_whiteplusoof}, with a low frequency cut-off at $f_{\rm cut}$ (see Fig.\,\ref{img_ps}). These values are used both for the noise simulations, and for the noise prior (Eq.~\ref{ps_teo}). The noise prior uses the same parameters of the injected noise simulation, except for the knee frequency in intensity, which takes a relaxed value of $40$\,Hz instead of $20$\,Hz (see text for details). \label{Table_noise_sim}}
\end{table}


\section{QUIJOTE-MFI map-making} \label{sec_QJT_mapmaking} 
The \namecode{} map-making code has been implemented to construct the maps of the MFI  instrument of the QUIJOTE experiment. 
%
In this section, first we briefly summarize the MFI instrumental response (Sec.~\ref{sec_instru_model}), we describe the QUIJOTE-MFI pointing matrix (Sec.~\ref{sec_proj_mat}), we derive the analytical equations for the \textit{I}, \textit{Q} and \textit{U} map (Sec.~\ref{sec_map}), and we set the noise priors  (Sec.~\ref{sec_cov} and \ref{solve_base}). Finally, in Sec.~\ref{sec_fitfunc} we describe the template function fitting at the map-making level.


\subsection{QUIJOTE-MFI instrumental response} \label{sec_instru_model}
The MFI has four feedhorns. Each horn has two frequency outputs, each of $2$\,GHz bandwidth. Horns number 1 and 3 are centered at $11$ and $13$\,GHz, and the horns number 2 and 4 at $17$ and $19$\,GHz. Each frequency of a given horn has four output channels that we call $\textbf{V}_{1},\,\textbf{V}_{2},\,\textbf{V}_{3},\,\textbf{V}_{4}$, which simultaneously observe the same point on the sky. The channels are grouped in two pairs. The channels that form the first pair,  $(\textbf{V}_{1},\,\textbf{V}_{2})$, present a relative correlation in their \oof noise component, so they are labelled as "correlated" channels. The second pair, $(\textbf{V}_{3},\textbf{V}_{4})$, did not have a relative correlation in the initial MFI instrumental setup, but the correlation was implemented at a later stage. For practical reasons, these channels are labelled as "uncorrelated". 
In total, there are 32 output channels. Finally, the QUIJOTE-MFI uses in-line polar modulators that modulate the polarization signal by four times the encoder angle. A complete description of the instrument can be found at \cite{MFIstatus12}. 

For a given horn and frequency band, each one of the four MFI channel provides a combination of intensity \textit{I} and linear polarization, the Stokes parameters \textit{Q} and \textit{U}, as:
\begin{align}
    \textbf{V}_{i} = \frac{1}{2}(I-(-1)^i &(\textbf{A}\sin(2\boldsymbol{\phi})Q+\textbf{B}\cos(2\boldsymbol{\phi})U+\\
    &+\textbf{C}\cos(2\boldsymbol{\phi})Q+\textbf{D}\sin(2\boldsymbol{\phi})U))+\textbf{n}_{i}
\end{align}
with $i=1,2,3,4$. $\boldsymbol{\phi}$ is related to the observed polarization direction with the relation $\boldsymbol{\phi}=2(\boldsymbol{\theta}-\boldsymbol{\theta}_0)+\boldsymbol{\phi}_p$, being $\boldsymbol{\theta}$, $\boldsymbol{\theta}_0$ and $\boldsymbol{\phi}_p$, respectively, the encoder, polar modulator reference and parallactic angles. $\textbf{A}$, $\textbf{B}$, $\textbf{C}$ and $\textbf{D}$ are in general vectors of the same length as the TOD, which can take values of either 0 or $\pm 1$, defining the equation of the detector response to the polarization signal, as a linear combination of \textit{Q} and \textit{U} modulated by sinusoidal functions of $2\boldsymbol{\phi}$. Usually, if \textit{Q} is modulated in the data by $\cos(2\boldsymbol{\phi} )$, then \textit{U} is modulated by $\sin(2\boldsymbol{\phi} )$, or vice versa, therefore we have not null values for only one of the parameters pairs:  even $(\textbf{A},\,\textbf{B})$ or $(\textbf{C},\,\textbf{D})$. Finally, in addition to the sky signal, the data contain a noise component $\textbf{n}$, that correlates as we mentioned above.  

A linear combination of these channels ($\textbf{V}_1$ with $\textbf{V}_2$ and $\textbf{V}_3$ with $\textbf{V}_4$) provides a measurement of the intensity and of the polarization of the sky signal. We can measure the intensity from the sum of pairs of channels, and the polarization from the difference. The polarization has the advantage that, when we make the difference of two correlated channels (for example $\textbf{V}_{1}-\textbf{V}_{2}$ or $\textbf{V}_{3}-\textbf{V}_{4}$), we cancel the correlated component of the noise. This can be written as:
\begin{align}
\label{tod_I}    \textbf y^{\rm I} & = \textbf{V}_{j}+\textbf{V}_{j+1}=I+\textbf n^{\rm I}&\\ 
\label{tod_QU}     \textbf y^{\rm P} & =  \textbf{V}_{j}-\textbf{V}_{j+1}=\textbf{A}\sin(2\boldsymbol{\phi})Q+\textbf{B}\cos(2\boldsymbol{\phi})U+&\\ 
             & \hspace{50pt}
             +\textbf{C}\cos(2\boldsymbol{\phi})Q+\textbf{D}\sin(2\boldsymbol{\phi})U+\textbf n^{\rm P}\notag
\end{align}
with $j=1,3$, where $\textbf y^{\rm I}$ and $\textbf y^{\rm P}$ are the TODs of the MFI, respectively for intensity and polarization, and $\textbf n^{\rm I}$ and $\textbf n^{\rm P}$ represent the noise component in the TOD, for intensity and polarization.  Thanks to the correlation between pairs of channels,  the $1/f$ component in the polarization noise TOD is much lower than that in intensity. However, if the gains of the channel pairs are not perfectly balanced, it leads to a residual $1/f$ in polarization too, which needs to be treated with the destriping technique.

In total, the MFI produces two sets of  $\textbf y^{\rm I}$  and  $\textbf y^{\rm P}$ TODs for each horn and frequency,  respectively from $(\textbf{V}_{1},\,\textbf{V}_{2})$ and $(\textbf{V}_{3},\textbf{V}_{4})$, allowing us to construct four independent maps of \textit{I}, \textit{Q} and \textit{U},  at each frequency.

The data sampling rate of the MFI is $1$\,ms, but the raw data are subsequently binned with a sampling rate of $t_{\rm s}=40$\,ms. The variance of the data in one $40$\,ms bin, named $\sigma^2$, is representative of the white noise level of the binned samples, if we assume that at the scale of one time bin the \oof{} drifts do not contribute to the noise variance. 
 We define the weights of the binned TOD elements as:
\begin{equation}
  w_i = \frac{1}{\sigma_i^2}.
  \label{weights}
\end{equation}
with $i=1,...,n_{\rm t}$. More details can be found in G{\'e}nova-Santos et al. (in prep.), describing the pipeline of QUIJOTE-MFI data.


\subsection{QUIJOTE pointing matrix}\label{sec_proj_mat}

Starting with the QUIJOTE-MFI instrumental response (Sec.~\ref{sec_instru_model}), we can represent a QUIJOTE-MFI TOD with and intensity and polarization part, using the notation:
\begin{equation}
\textbf y = \left(
\begin{array}{c}
     \textbf y ^{\rm I}\\
     \textbf y ^{\rm P}
\end{array} \right)\label{y}
\end{equation} 
where the vectors $\textbf y ^{\rm I}$ and $\textbf y ^{\rm P}$ of size $n_{\rm t}$ are defined in Eq.~\ref{tod_I} and \ref{tod_QU}. Similarly, the noise TOD can be written as:
\begin{equation}
\textbf n = \left(
\begin{array}{c}
     \textbf n ^{\rm I} \\
     \textbf n ^{\rm P} 
\end{array} \right)
\end{equation}
where $\textbf n^{\rm I}$ is the noise in the intensity TOD and $\textbf n^{\rm P}$ is the noise in the polarization one, both with size $n_{\rm t}$.
The sky map, in this notation, can be written as:
\begin{equation}
    \textbf m_{\rm sky} = \left(
\begin{array}{c}
     \textbf m ^{\rm I} \\
     \textbf m ^{\rm P}
\end{array} \right)
    = \left(
\begin{array}{c}
      I \\
     \left(
\begin{array}{c}
      Q \\
      U \end{array} \right) \\
\end{array} \right)
\end{equation}
 where we split again the intensity and polarization part. Let us define now the pointing matrix $\textbf{P}$ for QUIJOTE-MFI, which is a $(2n_{\rm t} \times 3n_{\rm p})$ elements matrix:
\begin{align} 
\textbf P &= \left(\begin{array}{ccc}
\textbf P^{\rm I}    & 0   \\  
0    & \textbf P^{\rm P}
\end{array}\right) 
\end{align}
being $\textbf P^{\rm I}$ a block matrix active on the intensity and $\textbf P^{\rm P}$ the block matrix of polarization. Given the QUIJOTE-MFI instrumental response (Sec.~\ref{sec_instru_model}), if pixel $j$ was observed at time $i$, the $_{ij}$ block of the pointing matrix \textbf P is:
\begin{align} 
\label{P_ij}
\textbf{P}_{ij} &= 
\left(\begin{array}{ccc}
 P^{\rm I}_{ij}    & 0   \\  
0    &  \textbf{P}^{\rm P}_{ij}
\end{array}\right) = \\
\notag
& = \left(\begin{array}{ccc}
1    & 0 & 0  \\  
0    & A_{i}\sin(2\phi_{i})+C_j\cos(2\phi_{i}) & B_{i}\cos(2\phi_{i})+D_{i}\sin(2\phi_{i})
\end{array}\right)
\end{align}
and it vanishes otherwise. The $P_{00}$ element interacts with the \textit{I} map to project the intensity in the TOD, while the elements $P_{11}$ and $P_{12}$ combine the \textit{Q} and \textit{U} maps into the polarization TOD, according to the instrumental response equations. If we develop now the TOD equation starting from Eq.~\ref{y=s+n}, and using the definitions in Eq.~\ref{y}-\ref{P_ij}, we obtain the instrumental response of Eq.~\ref{tod_I} and \ref{tod_QU}, showing the logic in the definition of the pointing matrix \textbf{P}. 

\subsection{\textit{IQU} analytical solution} \label{sec_map}

The \textit{I}, \textit{Q} and \textit{U} maps can be obtained by solving the map binning equation (Eq.~\ref{mout}), which applies the matrix $\textbf{M}^{-1}\textbf{P}^{T}\textbf{C}_{\rm w}^{-1}$ to the data subtracted by the baselines $(\textbf{y}-\textbf{F}\cdot \textbf a_{\rm out})$. Direct matrix multiplications are extremely expensive  computationally for any realistic data-sets, which involve a massive number of data. In order to avoid this problem, we derive an analytical solution for the maps of the three Stokes parameters, taking into account how the pointing matrix of QUIJOTE-MFI (Eq.\,\ref{P_ij}) projects them into the TOD. 

To get the intensity map we have to solve:
\begin{equation}
    \frac{\partial X}{\partial I}=0
\end{equation}
which gives:
\begin{equation}
    I_i = \frac{\sum_{j\in i} \frac{1}{\sigma_j^2}(y_j-
     \sum_{k=1}^{n_{\rm b}} F_{jk} a_k )}
    {\sum_{j\in i} \frac{1}{\sigma_j^2}}
    \,\,\,(i=1,...,n_{\rm p};\,j=1,...,n_{\rm t}) \label{I}
\end{equation}
with an associated variance:
\begin{equation}
    \sigma_{I_i}^2=\frac{1}{\sum_{j\in i} \frac{1}{\sigma_j^2}}\,\,\,(i=1,...,n_{\rm p};\,j=1,...,n_{\rm t})
    \label{Ierr}
\end{equation}
In these expressions, $i$ runs over the pixels, $j$ runs over the data samples with coordinates lying within that pixel. This tells us that the intensity in pixel $i$ is a weighted average of the baselines subtracted data that cross the pixel, where the weights are $1/\sigma_j^2$, as defined in  Eq.~\ref{weights}. 

Similarly, for the polarization, we have to solve:
\begin{equation}
\begin{cases}
    \frac{\partial X}{\partial Q}=0 \\\frac{\partial X}{\partial U}=0
\end{cases}
\end{equation}
that leads to the solutions:
\begin{equation}
   Q_i = \frac{a_if_i-d_ic_i}{a_ib_i-c_i^2} \label{Q}
\end{equation}
and 
\begin{equation}
   U_i = \frac{b_id_i-c_if_i}{a_ib_i-c_i^2} \label{U}
\end{equation}
with:
\begin{align}
a_i &= \sum_{j\in i} \frac{\left(B_j\cos(2\phi_j)+D_j\sin(2\phi_j)\right)^2 }{\sigma_j^2}\\
b_i &= \sum_{j\in i} \frac{\left(C_j\cos(2\phi_j)+A_j\sin(2\phi_j)\right)^2 }{\sigma_j^2}\\
c_i &= \sum_{j\in i} \frac{\left(B_j\cos(2\phi_j)+D_j\sin(2\phi_j)\right)\left(C_j\cos(2\phi_j)+A_j\sin(2\phi_j)\right)}{\sigma_j^2} \\
d_i &= \sum_{j\in i} \frac{\left(B_j\cos(2\phi_j)+D_j\sin(2\phi_j)\right)(y_j -  \sum_{k=1}^{n_{\rm b}}F_{jk}  a_k)}{\sigma_j^2} \\
f_i &= \sum_{j\in i} \frac{\left(C_j\cos(2\phi_j)+A_j\sin(2\phi_j)\right)(y_j -  \sum_{k=1}^{n_{\rm b}}F_{jk}  a_k)}{\sigma_j^2} \\
g_i &= a_i b_i -c_i^2
\end{align}
where $i$ runs over the pixels, and $j$ over the data. The coefficients $\textbf A,\,\textbf B,\,\textbf C$ and $\textbf D$ drive the combination of \textit{Q} and \textit{U} with $\sin(2\boldsymbol{\phi})$ and $\cos(2\boldsymbol{\phi})$, as explained in Sec.~\ref{sec_instru_model}. 
Finally, the variance maps of \textit{Q} and \textit{U}, and their covariance, can be computed as:
\begin{equation}
    \left(\sigma_Q^2, \sigma_U^2,  cov_{QU}\right)_i= \left(\frac{a_i}{g_i},\,\frac{b_i}{g_i}, -\frac{c_i}{g_i}\right)
    \label{QUerr}
\end{equation}
and a condition number ($r_{\rm cond}$) map, defined as the ratio of the two eigenvalues of the polarization block of the $\textbf{M}$ matrix \citep{Kurki-Suonio, Multidetector}, is computed as: 
\begin{equation}
    r_{\rm cond, i} = \frac{(a_i +b_i) + \sqrt{ (a_i -b_i)^2 + 4c_i^2  }}{(a_i +b_i) - \sqrt{ (a_i -b_i)^2 + 4c_i^2  }}.
    \label{eq:rcond}
\end{equation}
The $r_{\rm cond}$ map quantifies the goodness of the reconstruction of the Stokes \textit{Q} and \textit{U} maps.

It is worth noticing that the expressions in Eq.~\ref{Q} and~\ref{U} provide totally general solutions for the  polarization maps, in the sense that they can be applied to any instrument that measures a combination of \textit{Q} and \textit{U}, modulated by sinusoidal functions. In the case of QUIJOTE-MFI, the detector response changed several times during the multiple observing campaigns, due to upgrades and modifications of the instrument. This changed the way of combining \textit{Q} and \textit{U} into the TOD during different periods of observation. However, with the methodology that we described in this section, we can use a time varying combination of $\textbf A,\,\textbf B,\,\textbf C$ and $\textbf D$ to account for modifications of the instrumental configuration, and integrate all the data in one single \textit{Q} and \textit{U} map.

Finally, we have to specify that, in this implementation we do not account for the beam and pixel window function. Therefore, the result is a map of the sky convolved with the  beam window function of the experiment, and the pixel window function.


\subsection{Estimating the baselines with noise priors}\label{solve_base}

In Sections~\ref{sec:destr_algorithm} and~\ref{sec_cov} we derived the equations to estimate the baselines $\textbf a_{\rm out}$, which imply the solution of Eq.~\ref{aout} with a prior on the \oof noise, that is given by the covariance matrix of the baselines $\textbf{C}_{\text a}$ defined in Eq.~\ref{Ca_full}. However, direct  multiplication and inversion of matrices is too expensive computationally, and some approximations must be done here.

Equation \ref{aout} can be looked as the combination of three terms, which are: 
\begin{enumerate}
    \item ${\bf D} \cdot \textbf a_{\rm out} $
    \item $\textbf{C}_{\text a}^{-1}\cdot \textbf a_{\rm out} $ 
    \item $\textbf{F}^{T}\textbf{C}_{\rm w}^{-1}\textbf{Z}\textbf{y}$
\end{enumerate}
The problem consists in finding the vector $\textbf a_{\rm out}$ that satisfies the relation (i)+(ii)=(iii). To determine the solution for $\textbf a_{\rm out}$, we adopt the conjugate gradient method (CG), which allows us to move numerically in the baselines parameters space, towards the best $\textbf a_{\rm out}$ that satisfies Eq.\,\ref{aout}. With this aim, we have to compute (i) and (ii) for the set of numerically proposed solutions for  $\textbf a_{\rm out}$, in order to find the best match with (iii) within a given relative accuracy (which is set to $10^{-5}$ in \namecode{}). 

Terms (i) and (iii) can be determined analytically once the map binning equations are fixed (with Eq.~\ref{I},~\ref{Q},~\ref{U}). However, for (ii) we have to make one more approximation. It consists of computing the noise prior term $\textbf{C}_{\text a}^{-1}\cdot \textbf a$ in the Fourier space, where $\hat {\bf C}_{\text a}$ is diagonal (see Sec.~\ref{sec_cov}). Thanks to that, the inverse $\hat {\bf C}_{\text a}^{-1}$ is the straight scalar inversion of the diagonal elements of $\hat {\bf C}_{\text a}$, and we can compute the product $\hat{\bf C}_{\rm a}^{-1} \cdot \hat{\textbf a}$ as the element by element multiplication of the diagonal of $\hat{\textbf{C}}_{\text a}^{-1}$ (Eq.\,\ref{Ca_full}) with the Fourier transform of the baseline vector $\hat{\textbf a}$. Afterwards, we can project the result back to the real space, and obtain the multiplication $\textbf{C}_{\text a}^{-1}\cdot \textbf a$, which gives (ii). 


\subsection{Destriping QUIJOTE-MFI data with priors}
In the specific case of QUIJOTE-MFI, the TOD of intensity and polarization pass through the destriping step separately. In fact, as we mentioned in Section~\ref{sec_instru_model}, the noise in the intensity and in the polarization TOD are different. For the estimation of the baselines prior, we inject a theoretical power spectral density of the type of Eq.~\ref{ps_teo}, whose parameters are representative of the average noise properties of QUIJOTE-MFI. The parameters that we use are
listed in Table~\ref{Table_noise_sim}, and they are obtained as a result of the study of the typical noise properties of the MFI. The $f_{\rm k}$ and the $\gamma$ are input parameters of the map-making code, while the white noise level $\sigma$ is estimated from the data. $f_{\rm cut}=1/30$\,s$^{-1}$ is a fixed quantity specific for QUIJOTE, which corresponds to the frequency of one azimuth scan of $360$\,deg.  

In Fig.~\ref{img_aps} we show the diagonal of the baselines covariance matrix in Fourier space, as a function of the frequency, and for the aforementioned noise parameters. This plot represents the prior of the QUIJOTE-MFI noise, for intensity (orange line) and polarization (blue line).

\subsection{Fitting a template function} \label{sec_fitfunc}
We describe now the additional feature that we implemented in \namecode{}: the fitting of a template function at the TOD level, during the map-making step. With the same logic of cleaning the noise with baselines, we extend the destriper algorithm in order to fit the data with a specific template $\textbf{f}$ in the time domain. We can write the TOD as:
\begin{equation}
  \textbf{y}'= \textbf{y} + A \cdot \textbf{f}
  \label{eq:TOD_fitfunc}
\end{equation}
where we added to the data $\textbf{y}$ (Eq.\,\ref{y=fa+w+pm}) a template function $\textbf{f}$ with amplitude $A$. The technique to determine the amplitude $A$ can be seen as en extension of the usual destriping  presented in Sec.~\ref{sec:destr_algorithm}, where we added the $A \cdot \textbf{f}$ component in the model of the TOD, and where we apply the simplification of neglecting the noise priors term (${\bf C}_{\rm a}^{-1}=0$). We construct the chi-square, similarly to Eq.\,\ref{eq:X}, as:
\begin{equation}
     X=(\textbf{y}'-\textbf P \cdot \textbf m_{\rm sky}-\textbf F \cdot \textbf a-A \cdot \textbf{f})^{T}\textbf{C}_{\rm w}^{-1}(\textbf{y}'-\textbf P  \cdot \textbf m_{\rm sky}-\textbf F  \cdot \textbf a-A \cdot \textbf{f})+{\rm const}
      \label{eq:X_impl}
\end{equation}
The minimization with respect to $A$ of $X({\bf m}={\bf m}_{\rm out},\,{\bf a}={\bf a}_{\rm out})$, with ${\bf m}_{\rm out}$ and ${\bf a}_{\rm out}$ given by Eq.\,\ref{mout} and \ref{aout} (with ${\bf C}_{\rm a}^{-1}=0$), provides the equation for the fitted template amplitude:
\begin{equation}
  A_{\rm out}=  C^{-1} \textbf{f}^T \textbf{C}_{\rm w}^{-1}\textbf{Q}\cdot \textbf{y} 
  \label{A_out}
  \end{equation}
being $C$ a scalar number, and $\textbf{Q}$ a $(n_{\rm t} \times n_{\rm t})$ matrix defined as:
\begin{align}
 C&=\textbf{f}^T\textbf{C}_{\rm w}^{-1}\textbf{Q}\textbf{f} \\
 \textbf{Q}&=\textbf{Z}(\textbf{1}-\textbf{FD}^{-1}\textbf{F}^T\textbf{C}_{\rm w}^{-1}\textbf{Z})
\end{align}
Once $A_{\rm out}$ is determined, we construct the template subtracted TOD $\textbf{y}= \textbf{y}' - A_{\rm out} \cdot \textbf{f}$, and proceed with the usual destriping presented in Sec.~\ref{sec:destr_algorithm}.

We can analytically derive the uncertainty on $A_{\rm out}$ with a Fisher-matrix approach, assuming Gaussianity of the posterior distribution and that the noise is uncorrelated. It is given by:
\begin{equation}
\sigma_ {A_{\rm out}}= \frac{1}{\sqrt C } 
\label{errA}
\end{equation}
However, with the actual MFI intensity data containing large $1/f$ noise, the assumptions that we made to draw Eq. \ref{errA} are not necessarily valid, leading to an underestimated uncertainty on $A_{\rm out}$. A more realistic estimate for $\sigma_ {A_{\rm out}}$ can be obtained with Monte Carlo simulations, as we will show in Sec.~\ref{sec_fit_results}.

The template fitting method presented in this section can be used for multiple purposes, both in intensity and polarization, and with any kind of templates. We show in Section~\ref{sec_fit_results} two different applications of particular interest for microwave wavelength data: we fit the CMB dipole and an atmospheric component to the intensity. In addition, this fitting procedure could be easily generalized  to fit the templates in selected chunks of data instead that in the full data-set, or to fit several template functions simultaneously. For a number of $n_{\rm A}$ templates, ${\bf A}$ is a vector of $n_{\rm A}$ elements, ${\bf f}$ an $(n_{\rm t},\,n_{\rm A})$ template matrix, and ${\bf C}$ an $(n_{\rm A},\,n_{\rm A})$ matrix.


\section{The code} \label{sec_code}
The \namecode{} code has been developed for the map-making of the QUIJOTE experiment. It is a {\sc F90} implementation, with parallelization based on {\sc OpenMP}. \namecode{} is built in two main blocks: the first for the data reading (see Sec.~\ref{sec:data}-\ref{sec_template}), and the second for the construction of the baselines subtracted intensity and polarization maps (see Sec.~\ref{sec:destriping}, \ref{sec:binningIQU}). While the data reading is specific for the MFI, the block of map-making is totally general, and is organized in three steps: (i) the TOD template fitting, if required by the user (Sec.~\ref{sec_template}), (ii) the destriping (Sec.~\ref{sec:destriping}), and (iii) the projection of the cleaned data into the map (Sec.~\ref{sec:binningIQU}). In step (iii) we also build a map of the number of hits in each pixel ($n_{\rm hits}$), the error map, and the $r_{\rm cond}$ map (as described in Sec.~\ref{sec_map}).

In order to characterize the effect of the noise in the maps with simulations, the code is implemented with a noise generator (see Sec.~\ref{Noise}). This allows us to dynamically add to a simulated sky signal TOD a noise realization, with a power spectral density of the type of Eq.~\ref{ps_teo_whiteplusoof}.

\subsection{Data}\label{sec:data}

The \namecode{} map-making code is currently implemented for the QUIJOTE-MFI data, whose instrumental response is described in Sec.~\ref{sec_instru_model}. In order to construct the TOD of a selected MFI detector and frequency, we import the data of the MFI channel pairs (e.g., Horn 3, 11\,GHz, $\textbf{V}_{1}$ and $\textbf{V}_{2}$), and we combine them to construct the intensity or polarization TOD, as in the Eq.~\ref{tod_I} and~\ref{tod_QU}. The TODs of the full set of observations that we want to process are then stored in memory, together with their weights and pointing coordinates, in order to proceed with the construction of the map. 

It should be noticed that the simultaneous combination of a large amount of observations in one single map is very important. Indeed, it is recommended to have a large number of crossings on the same pixel with different scan orientations on the sky, in order to perform a precise determination of the baselines, and also of the \textit{Q} and \textit{U} Stokes parameters which need to be sampled with different orientations of the angle $\boldsymbol{\phi}$. With this aim, the  map-making code must be able to handle a large amount of data.

\namecode{} has been used to construct the QUIJOTE-MFI wide-survey maps. The full data-set of the QUIJOTE wide-survey consist of $\sim 11042$\,h of observation, amounting to $\sim 340$\,Gb. For the construction of one single map, we need to store in memory the data of one selected pair of channels, which corresponds to $\sim 340/16$\,Gb\,$\approx 21$\,Gb. One single core of a machine that is capable to store this amount of data in memory takes about 15-20\,hours to produce a map. 


\subsection{Noise simulations}\label{Noise}
In \namecode{}, we implemented the option to generate and add noise simulations to the TOD, simultaneously with the data reading. With this utility, from one single end-to-end simulation of the sky signal, we can produce a number of realizations of noisy maps, by  adding the noise on-the-fly.   To simulate the noise, we use a Fourier inversion technique of an  input noise power spectral density of the type as in Eq. \ref{ps_teo_whiteplusoof}. The noise generator has been used to produce the simulations with which we tested the template function fitting, which are presented in Sec.~\ref{sec_fit_results}.

\subsection{Gaps in the data}\label{sec:gaps}
The good data in a TOD are usually alternated with corrupted data,  which can be affected by RFI, instrumental problems, bad weather, etc. Moreover, the MFI adopts a calibration technique that injects signal from a calibration diode superimposed on top of the sky signal, for one second every thirty seconds, and these calibration data must also be discarded.  

To deal with the interruptions, we adopt the strategy of down-weighting the flagged data, as in \cite{Planck-mapmaking}, by assigning to the bad data zero weights ($w_i=\sigma_i^{-2}=0$, see Eq.\,\ref{weights}), which is equivalent to setting an infinite variance.

 \subsection{Template functions} \label{sec_template}
As we mentioned in Section~\ref{sec_fitfunc},  we implemented an algorithm to fit for a template function at the TOD level, during the map-making step. In order to apply this technique, we need to define a TOD of the template.  The code has the capability to generate a template from a totally general map that is delivered by the user, by  projecting this map  into a template with TOD format. In addition, two special templates are internally generated by \namecode{}: a template of the atmosphere (see Sec.~\ref{sec_atmo}) and of the CMB dipole (see Sec.~\ref{sec_dipole}). The results of the fit of these two templates are reported in Sec.~\ref{sec_fit_results}.

The template fitting is applied to the TOD before  destriping, independently in intensity and polarization for QUIJOTE-MFI data.\footnote{If the template fitting is applied to the intensity, there is no effect in the polarization maps, and the other way around.} Once the amplitude of the template $A_{\rm out}$ is estimated, we subtract the $A_{\rm out} \textbf{f}$ component from the data (see Eq.\,\ref{eq:TOD_fitfunc}), which then passes through the destriping step, and is finally binned into the \textit{I}, \textit{Q} and \textit{U} maps.  

We describe now the two templates that are internally implemented into the code: a stable, plane-parallel atmosphere, and the CMB solar plus orbital dipole. However, \namecode{} can be extended with other templates of typical contaminants of ground-based observations, like ground pickup of RFI contamination.

\subsubsection{Atmosphere} \label{sec_atmo}
A not perfect alignment of the azimuth axis of the telescope mount with the zenith direction can introduce in the data a modulated component of atmospheric emission. Indeed, when the telescope performs azimuth scans at a fixed elevation (as it is the case for QUIJOTE - see Sec.~\ref{Scanning}), if the azimuth axis is not perfectly aligned to the zenith, the effective (true) elevation of the scan is not perfectly constant, and the line of sight crosses varying air masses. This leads to a measurement of a modulated atmospheric emission given by:
\begin{equation}
    \textbf{f}_{\mathrm{atm}}=\frac{T_{\mathrm{atm}}}{\sin(\textbf{el})} \label{eq:sinel}
\end{equation}
where $\textbf{el}$ is a TOD of the real elevation of the observation, and $T_{\mathrm{atm}}$, which is the amplitude of the modulation, is the antenna temperature of the atmosphere at the Zenith\footnote{$T_{\mathrm{atm}}=T_{\mathrm{atm},0}\cdot(1-e^{-\tau}) \sim T_{\mathrm{atm},0}\tau$, where $T_{\mathrm{atm},0}$ is the temperature of the atmosphere, and $\tau$ is the atmospheric opacity.}, at a determined frequency.  Here we assume a static, plane-parallel atmosphere\footnote{The atmosphere in the plane-parallel approximation is assumed to be composed by consecutive two-dimensional parallel planes. In this approximation, the angle between the direction of propagation of light across the atmosphere and the normal to the planes is constant.}, which is a sufficiently good approximation for our purposes\footnote{Taking as a reference the exact equation of atmospheric emission given in \cite{Kasten1989}, it can be noticed that the correction to the approximated Eq.\,\ref{eq:sinel}, at $el=30^{\circ}$, is just 0.3\%, and is lower at higher elevations.}.

In QUIJOTE the degree of misalignment of the azimuth axis is less than $0.05^\circ$, which for a typical precipitable water vapour (PWV) value of $3$\,mm, corresponding to a temperature amplitude of $T_{\mathrm{atm}}=1.8$\,K at 11\,GHz \citep{Pardo2001}, produces temperature fluctuations of about 2\,mK, at elevation of $60^{\circ}$. 

In the \namecode{} code, we generate the template of the atmosphere with the function:
\begin{equation}
    \textbf{f}=\frac{1}{\sin(\textbf{el})}
    \label{template_atmo}
\end{equation}
using the true value of the elevation $\textbf{el}$ as derived from the QUIJOTE pointing model, and we recover the amplitude $A=T_{\mathrm{atm}}$, which is the average temperature of the atmosphere at the zenith during the observations (normalized by units of Kelvin).

\subsubsection{CMB dipole}  \label{sec_dipole}
The Sun's motion with respect to the CMB reference frame produces, via Doppler effect, a solar dipole anisotropy of the CMB radiation, with an amplitude of $\sim 3$\,mK. In addition, the Earth's motion around the Sun introduces an extra dipole anisotropy, the so-called orbital dipole, with a smaller amplitude than the solar dipole, of $\sim 200\,\mu$K. 

We call $\textbf{v}=\textbf{v}_{\rm s}+\textbf{v}_{\rm e}$ the combination of the velocity vector of the Sun $\textbf{v}_{\rm s}$ with respect to the CMB reference frame, and the velocity of the Earth  $\textbf{v}_{\rm e}$ with respect to the Sun, in order to model jointly the solar and orbital CMB dipole. For the solar dipole velocity $\textbf{v}_{\rm s}$ we used the HFI 2018 dipole from \citealt{Planck2018HFI}, combined with the CMB temperature  $T_{\mathrm{CMB}}=2.72548$\,K from \citealt{Fixsen2009} to convert temperature to velocity (back and forth), while the Earth's velocity $\textbf{v}_{\rm e}$ is predicted from the JPL Horizons ephemeris\footnote{\href{https://ssd.jpl.nasa.gov/horizons.cgi}{https://ssd.jpl.nasa.gov/horizons.cgi}}.
 In this framework, the CMB solar plus orbital dipole  can be precisely predicted in time, for any direction of observation, and we can build a time ordered template of the predicted total CMB dipole (Eq.\,\ref{eq_dipole}). 

In the first order regime, where the ratio between the module of the velocity $v=|\textbf{v}_{\rm s}+\textbf{v}_{\rm e}|$ and the speed of light $c$ is small ($\beta=v/c\approx 10^{-3} \ll 1$), the CMB dipole is given by (\citealt{peebles1968}):
\begin{equation}
    \textbf{f}_{\rm dip} \approx  T_{\mathrm{CMB}}\boldsymbol{\beta} \cos(\boldsymbol{\theta}')
    \label{eq_dipole}
\end{equation}
where $T_{\mathrm{CMB}}\boldsymbol{\beta}$ is the instantaneous amplitude of the CMB solar plus orbital dipole, and  $\boldsymbol{\theta}'$ is the angle between the velocity of the observer and the direction of observation, as a function of time. 
Eq.~\ref{eq_dipole} can be used as a template function of the CMB dipole in the time ordered domain. A fit of $\textbf{f}=\textbf{f}_{\rm dip}$ can be done during the map-making procedure, by applying the technique that we described in Section~\ref{sec_fitfunc}. The resulting amplitude is expected to be $A_{\rm out}=1$.

\subsection{Destriping}\label{sec:destriping}
Destriping consists in solving Eq.~\ref{aout}, whose solution, the baselines vector  $\textbf{a}_{\rm out}$, provides a model of the \oof{}noise to be subtracted from the TOD (see Sec.~\ref{sec_destriper} and \ref{solve_base}). In \namecode{}, we implemented this step with three options, where we allow the user to select which type of priors on the baselines he/she wants to apply. The options are: 
\begin{enumerate}
\item No prior on the baselines (i.e., $\textbf{C}_{\text a}^{-1} = 0$).
\item Approximation with diagonal baselines covariance matrix, $C_{\text a,{ij}} = \delta_{ij} \left<\sigma_i\sigma_j\right>/n_{\rm b} $ where $\sigma_i^2$ is the rms of the noise within the baseline $i$.
\item Full prior of the baselines, where $\textbf{C}_{\text a}$ is given by Eq.~\ref{Ca_full}, and is applied with Fourier techniques, as described in Section~\ref{solve_base}.
\end{enumerate}

Option (i) was used to construct the maps presented in \cite{W44} and \cite{Taurus}, option (ii) has only been used during the testing and validation phase of the code, while (iii) is the default option to construct the wide-survey maps (Rubi{\~n}o-Mart{\'{\i}}n et al. in prep).  
In option (iii), we assume stationary \oof noise properties across the full data-set, with a \oof power spectral density given by Eq.~\ref{ps_teo}. The sampling frequency $f_{\rm s}$ and the white noise level $\sigma$ are estimated from the data, while the knee frequency $f_{\rm k}$ and $\gamma$ are selected as input values by the user, and are assumed to be stationary. 
%

However, in the real QUIJOTE data, as in any other ground-based experiment, we do not have a perfectly stable \oof across all the observations, especially in intensity (Rubi{\~n}o-Mart{\'{\i}}n et al. in prep). In order to test the effect of a prior which is slightly different from the actual noise in the data, we produced intensity maps with a knee frequency prior $f_{\rm k}=40$\,Hz (see Tab.~\ref{Table_noise_sim} for a summary of the noise parameters), which is different from the actual value which was adopted to generate the simulations ($f_{\rm k}=20$\,Hz).  
%
The realistic noise simulations that are presented in this paper (Sec.~\ref{Sec_simu}), as well as the real wide-survey maps, are constructed using option (iii), with the prior parameters in Tab.~\ref{Table_noise_sim}. 

A detailed study for the selection of the baseline length was performed by \cite{alba}, with a work based on QUIJOTE-MFI raster scans observations. They showed that $t_{\rm b}=2.5$\,s provide optimal noise performance, therefore we select this baseline length also for the construction of wide-survey maps. 
Below, we test that this choice also produces (nearly) optimal results for the wide-survey observing mode.
However, in general, $t_{\rm b}$ is an arbitrary parameter that can be selected by the user when running \namecode{}. The calculation of the  baselines is performed  using the full data-set, in order to gain as much information as possible from different crossing of the same pixel.
 
\subsection{Projecting into the \textit{I}, \textit{Q}, and \textit{U} maps}\label{sec:binningIQU}
The projection of the baselines subtracted TOD $(\textbf y - \textbf{F} \cdot \textbf{a}_{\rm out})$ into a map is made by solving Eq.~\ref{mout}, as described in Section~\ref{sec_map}. The pixel size can be selected by the user, who is asked for the desired resolution in terms of the $N_{\rm side}$ HEALPix parameter \citep{Gorski2005}. We used $N_{\rm side}=512$ to construct the QUIJOTE-MFI maps, which corresponds to an angular resolution of $\sim6.9$\,arcmin. 

In \namecode{}, the map binning step is done separately for intensity and polarization: we construct the maps by applying the analytical solution formulated, for the intensity, in Eq.\,\ref{I}, and for the \textit{Q} and \textit{U} Stokes parameters, in Eq.\,\ref{Q} and \ref{U}. The variance maps of \textit{I}, \textit{Q}, and \textit{U} are constructed, respectively, with Eq.\,\ref{Ierr} and \ref{QUerr}. Finally, in the process of constructing the maps, we sum the number of data samples that hit each pixel, and we produce the so-called $n_{\rm hit}$ map. In the case of polarization, the \textit{Q} and \textit{U} maps are stored only in pixels where the number of hits is greater or equal to 3 ($n_{\rm hit} \geq 3$), in order to ensure the minimal number of crossings of the pixel necessary to determine the \textit{Q} and \textit{U} parameters from one TOD.\footnote{One polarization TOD is a function of two parameters, \textit{Q} and \textit{U}, combined with sinusoidal functions (see Eq.\,\ref{tod_QU}). Two observations of the same pixel ($n_{\rm hit}=2$) with two different parallactic angles would allow us to reconstruct \textit{Q} and \textit{U}, but we decided to be more conservative using a minimum $n_{\rm hit}=3$.}  In the case of polarization, we also build the $r_{\rm cond}$ map formulated in Eq.\ref{eq:rcond}, which quantifies the goodness of the \textit{Q} and \textit{U} reconstruction, given the variety of orientations of the angle $\boldsymbol{\phi}$ in each pixel. In addition, pixels with $r_{\rm cond}>3$ are excluded from the maps in the post processing stage.
 
 Since the MFI provides two TODs for each horn and frequency of the instrument (from the pairs of channels $(\textbf{V}_{1}\pm\,\textbf{V}_{2})$ and $(\textbf{V}_{3}\pm\,\textbf{V}_{4})$, where the sum is for the intensity and the difference for polarization, as explained in Sec.~\ref{sec_instru_model}), a set of 16 \textit{I}, \textit{Q} and \textit{U} maps is produced, with separate runs of the map-making code for intensity and polarization. The two maps from the same horn and frequency (e.g., Horn 3, 11\,GHz, from $(\textbf{V}_{1}\pm\,\textbf{V}_{2})$ and $(\textbf{V}_{3}\pm\,\textbf{V}_{4})$) are combined a posteriori with a weighted average (as in Eq.\,\ref{eq:wei_avg_map} and \ref{eq:wei_avg_wei}),  giving as a final result a set of 8 \textit{I}, \textit{Q} and \textit{U} maps, one for each of the two frequencies of the four horns. 
 
 

\section{Simulations}\label{Sec_simu}
The \namecode{} map-making code is tested and validated with simulations. Although the code is general and can be used to map any kind of observation (wide-survey or raster scans), in this paper we used a realistic simulation of the QUIJOTE-MFI wide-survey (Rubi{\~n}o-Mart{\'{\i}}n et al. in prep). The simulated TODs contain sky signal following the equations of the instrumental response of the MFI (Eq.~\ref{tod_I} and~\ref{tod_QU}), and three configurations of noise: no-noise, white only noise, and realistic white plus correlated \oof noise. Analogously to what is done to the real data, every $30$\,s we subtracted a median value from the TOD, with the goal of filtering to first order the \oof{} noise in one azimuth scan of $360$\,deg. The template fitting is applied to the TODs before subtracting the medians.

In the next sub-sections we describe the scanning strategy (Sec.~\ref{Scanning}), the sky signal (Sec.~\ref{signalsim}), and the noise (Sec.~\ref{noisesim}) that are used for the simulations.

\subsection{Wide survey scanning strategy} \label{Scanning}
The scanning strategy of the QUIJOTE-MFI wide-survey observations consists of continuous spins of the telescope at a constant elevation, with long term elevation re-pointing, at $el = [30,\,35,\,40,\,50,\,60,\,65,\,70,\,75,\,80]$\,deg. The scanning speed is $12$\,deg/s, and one azimuth scan of $360$\,deg is completed every 30\,s. After $24$\,h of observation we cover the full sky observable from the Teide Observatory (Tenerife, Spain, 2400\,m a.s.l.), with a coverage in declination that depends on the elevation of the observation. 

The effective time of observation is $\sim 11042$\,h, 
across four different instrumental setups (or periods). Elevation $el=30^{\circ}$ gives the widest declination coverage, which is $dec \in [-35, 90] \deg$, corresponding to a sky coverage of $78\%$. A complete description of the data-set can be found in Rubi{\~n}o-Mart{\'{\i}}n et al. (in prep.).

\begin{figure*}
\centering
    \includegraphics[width=0.32\textwidth]{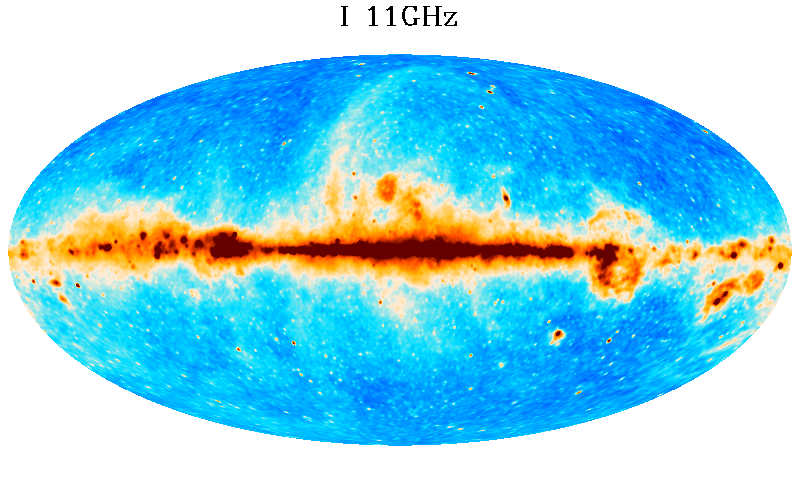} \includegraphics[width=0.32\textwidth]{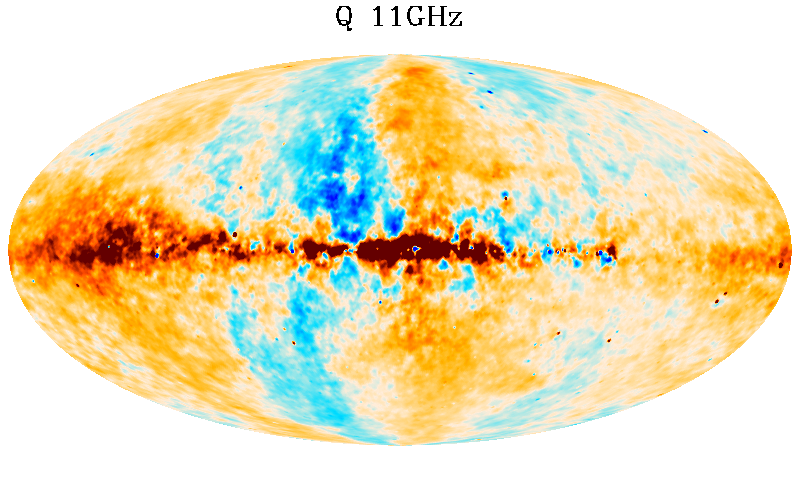} \includegraphics[width=0.32\textwidth]{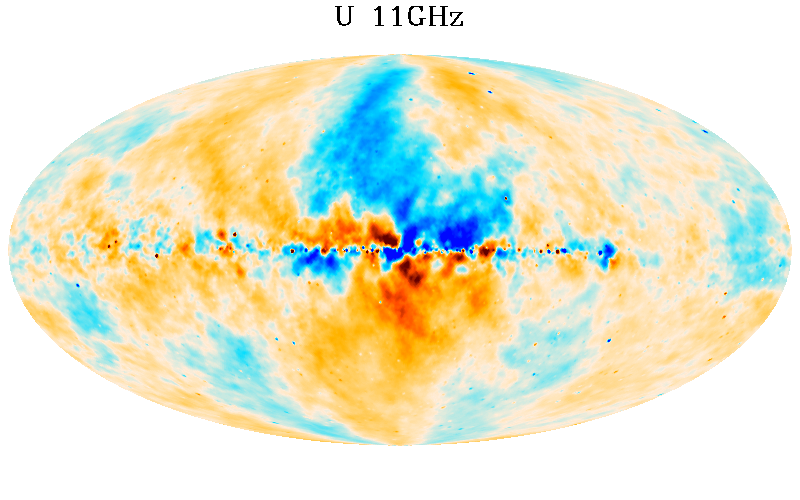}
    \includegraphics[width=0.32\textwidth]{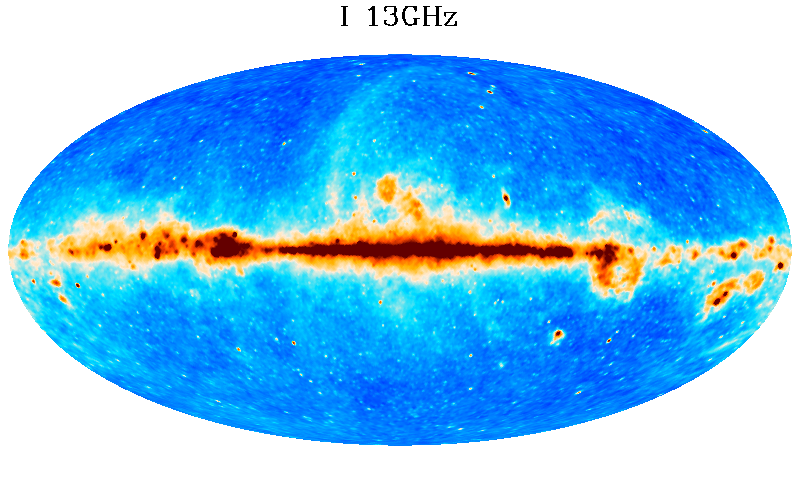} \includegraphics[width=0.32\textwidth]{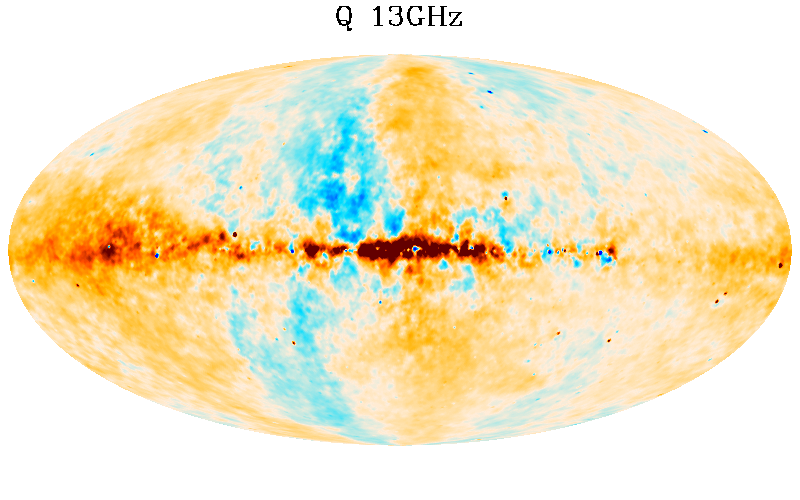} \includegraphics[width=0.32\textwidth]{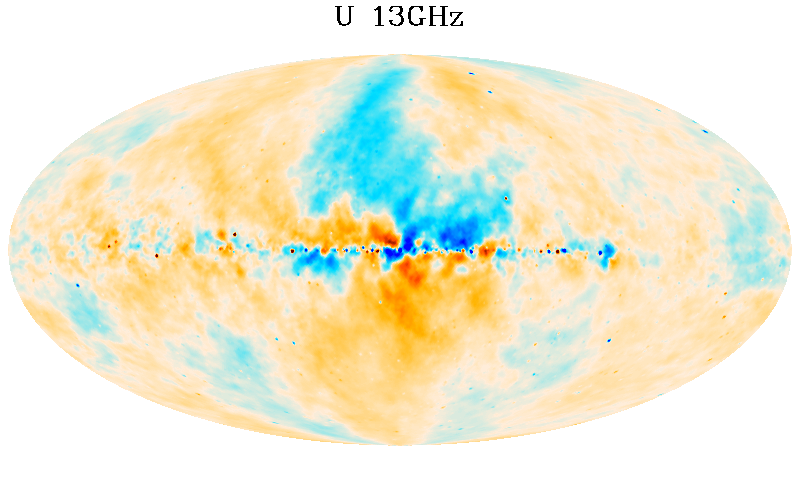}
    \includegraphics[width=0.32\textwidth]{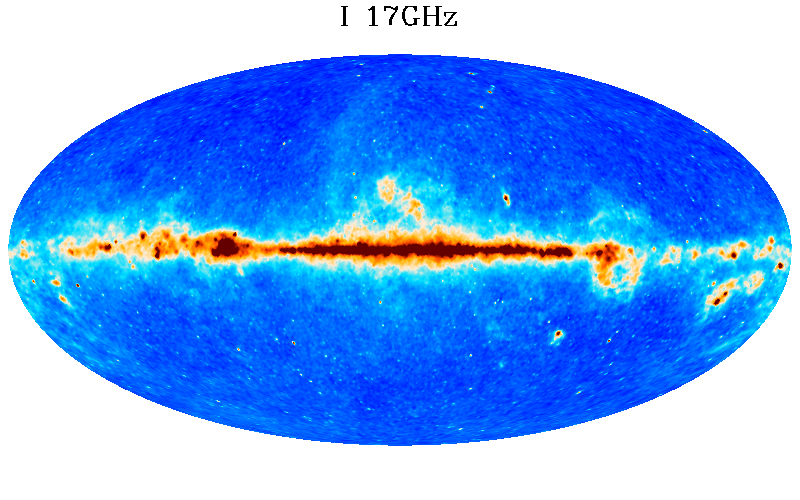} \includegraphics[width=0.32\textwidth]{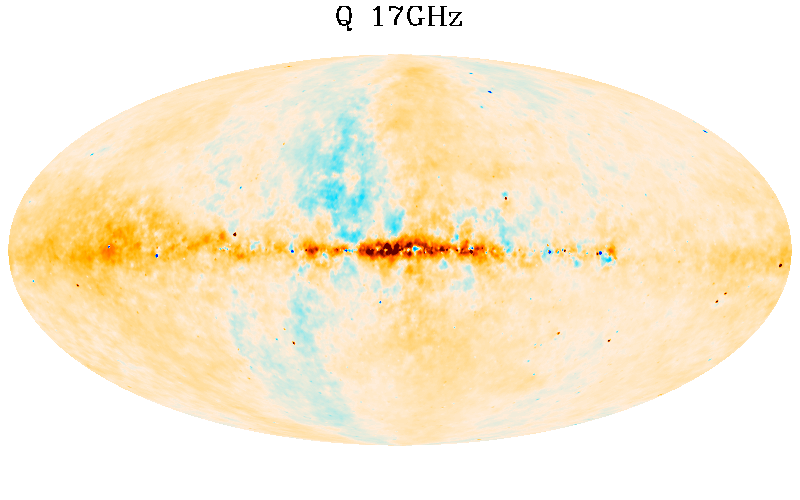} \includegraphics[width=0.32\textwidth]{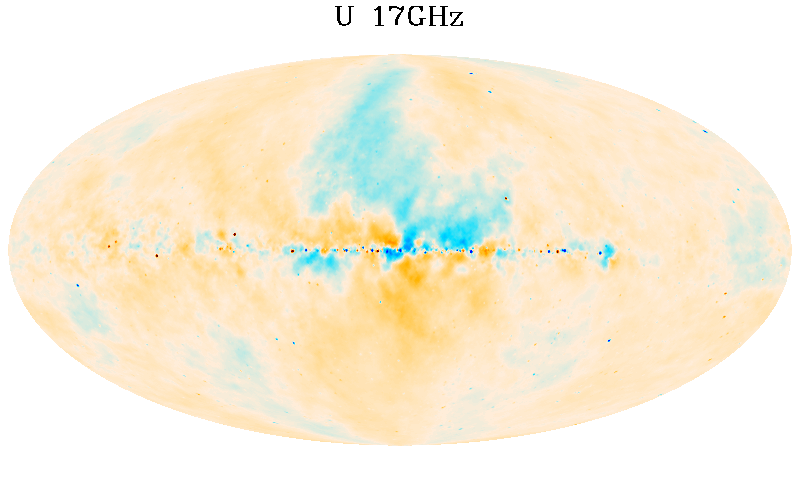}
    \includegraphics[width=0.32\textwidth]{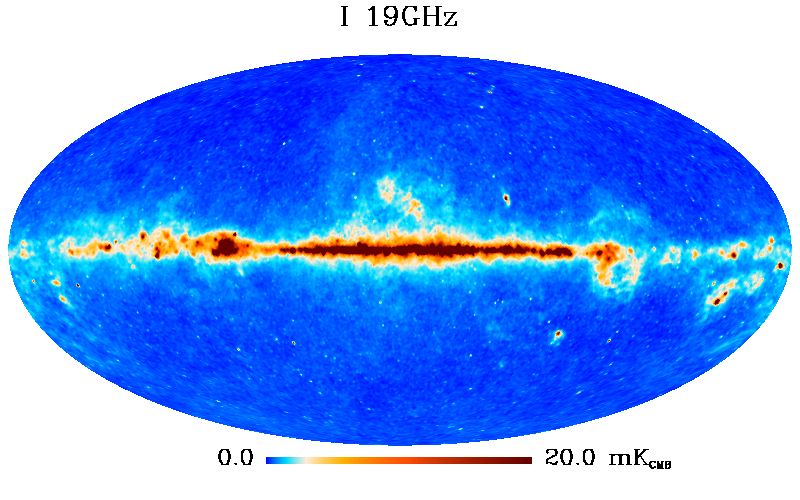} \includegraphics[width=0.32\textwidth]{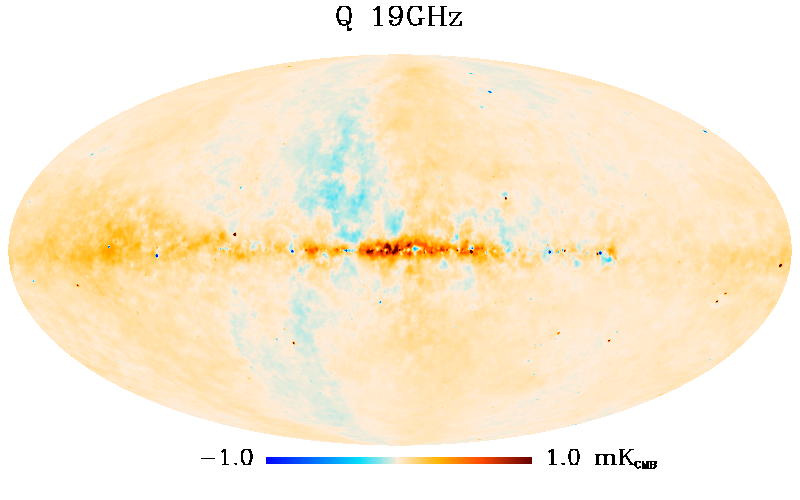} \includegraphics[width=0.32\textwidth]{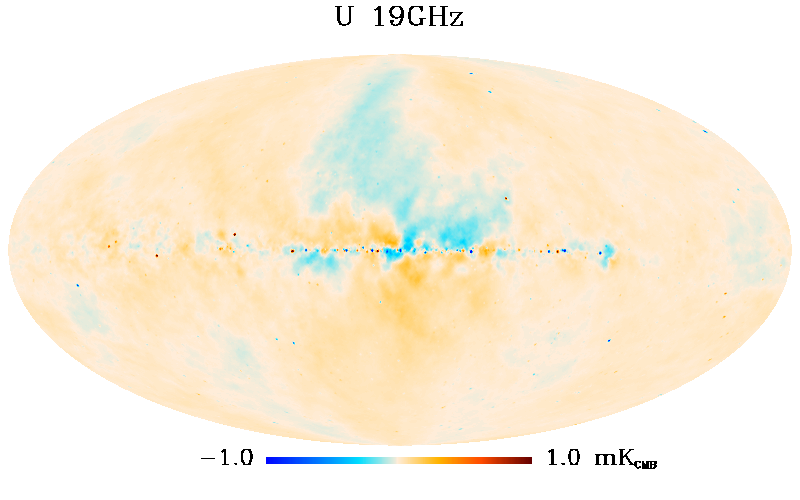}
\caption{Simulated sky signal \textit{I}-\textit{Q}-\textit{U} full sky maps, in galactic coordinates. The rows show, in order, the maps at 11, 13, 17 and 19\,GHz. The columns show, in order, the maps of Stokes \textit{I}, \textit{Q}, and \textit{U} parameters. }
\label{fig_sims}
\end{figure*}
\subsection{Simulated sky signal}
\label{signalsim}

We simulated the sky signal by projecting the intensity and polarization maps of a simulated sky into the TODs, according to the MFI instrumental response equations  described in Sec.~\ref{sec_instru_model}. The sky signal simulated maps that we use in this work were developed in the context of the RADIOFOREGROUNDS project.\footnote{\url{http://www.radioforegrounds.eu/}}

The sky simulations at the four MFI frequencies  contain foregrounds from the Planck FFP10 sky model (\citealt{Planck2018,Planck2018HFI}), a CMB realization from the Planck 2015 best-fit cosmology (\citealt{Planck2015}) with tensor-to-scalar ratio $r=0$, and a realistic CMB solar plus orbital dipole (from Eq.~\ref{eq_dipole}). The simulated sky input maps are shown in Fig.~\ref{fig_sims} (without the CMB dipole component). Finally, we added a template of the atmosphere (from Eq.~\ref{eq:sinel}, with $T_{\mathrm{atm}}=1$\,K), only when we wanted to test the fitting of this component.
The simulations are convolved at the native angular resolution of the experiment, approximated with a Gaussian beam with full width half maximum (FWHM) of $0.85$\,deg at $11$ and $13$\,GHz, and $0.63$\,deg at $17$ and $19$\,GHz. 

These maps are used as a reference sky to produce synthetic TOD vectors. In this step, the pointing coordinates, the weights and the flags are extracted from the TODs of the corresponding real wide-survey observations. 

\subsection{Noise}\label{noisesim}
We considered three possible configurations for the simulated noise: no-noise, white noise only, and realistic white plus correlated \oof noise. The noise is added to the simulated input sky, at the TOD level. In the white only case, the noise is produced as a random Gaussian realization with a variance given by the inverse weight of the data samples (Eq.~\ref{weights}). In this way, the noise mimics the actual noise variance structure in the real data. 
In the case of realistic white plus \oof noise, we use inverse Fourier techniques based on a power spectral density given by Eq.~\ref{ps_teo_whiteplusoof}, computed with the realistic noise parameters of the MFI reported in Table~\ref{Table_noise_sim}. The \oof parameters are assumed to be stable in time, and they are the same for all the MFI channels. 

\section{Results} 
\label{sec_results}

\begin{figure*}
\centering
    \includegraphics[width=0.24\textwidth]{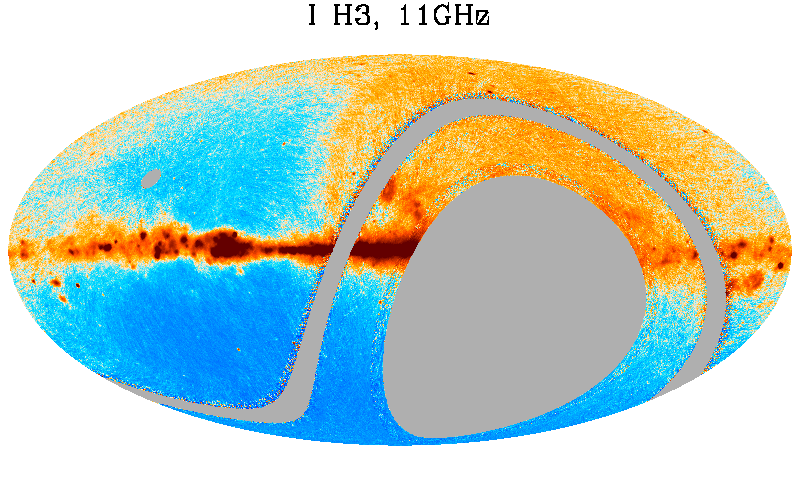} \includegraphics[width=0.24\textwidth]{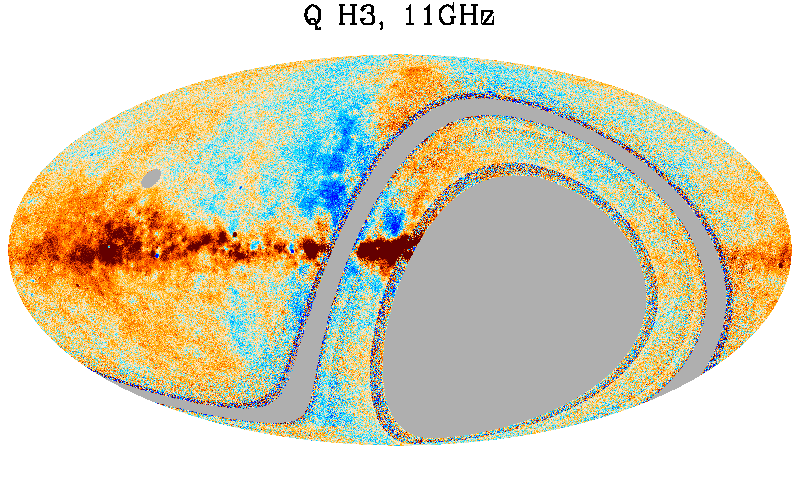} \includegraphics[width=0.24\textwidth]{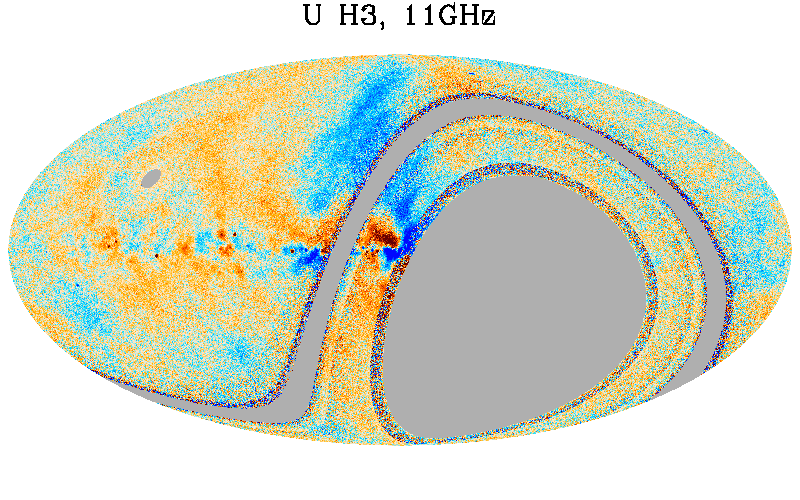}
    \includegraphics[width=0.24\textwidth]{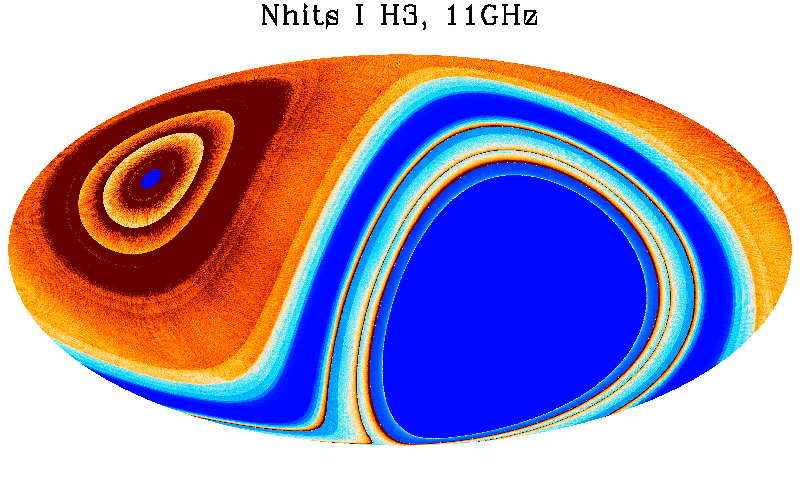}
    
    \includegraphics[width=0.24\textwidth]{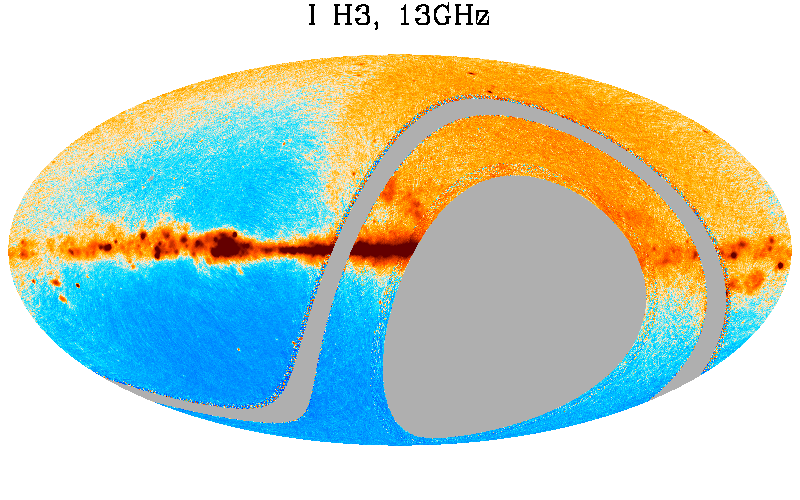} \includegraphics[width=0.24\textwidth]{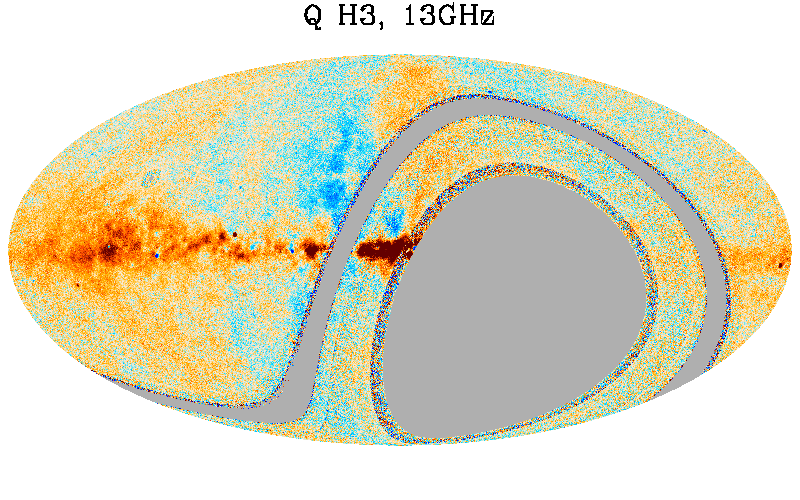} \includegraphics[width=0.24\textwidth]{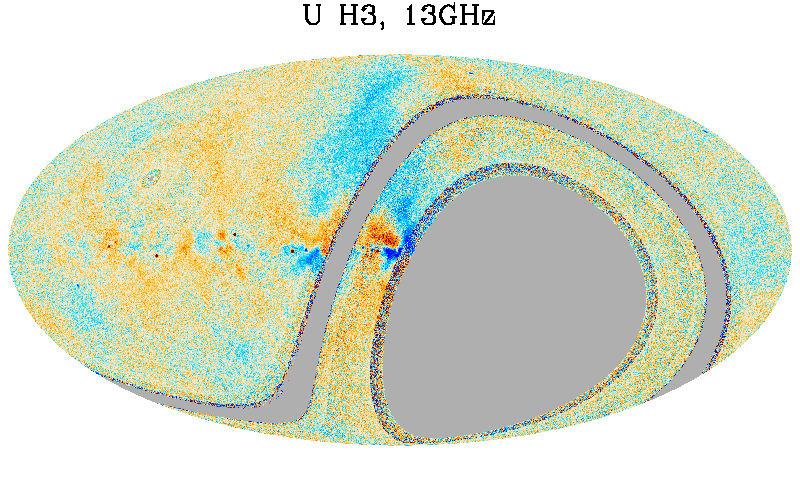}
    \includegraphics[width=0.24\textwidth]{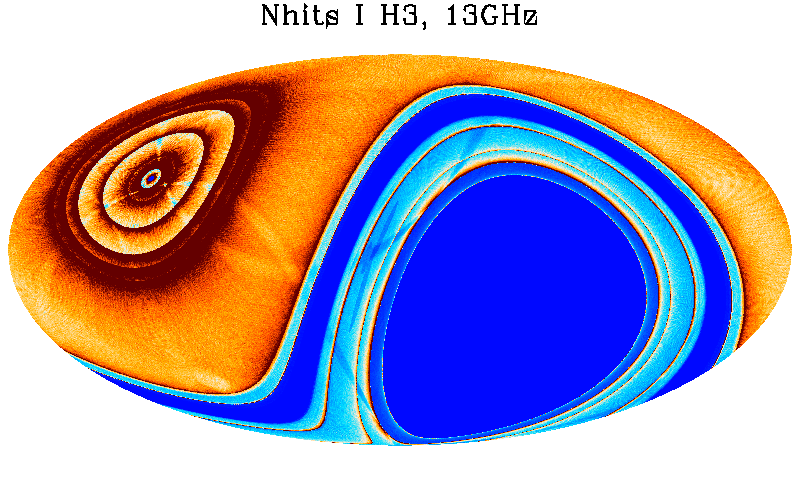}
    
    \includegraphics[width=0.24\textwidth]{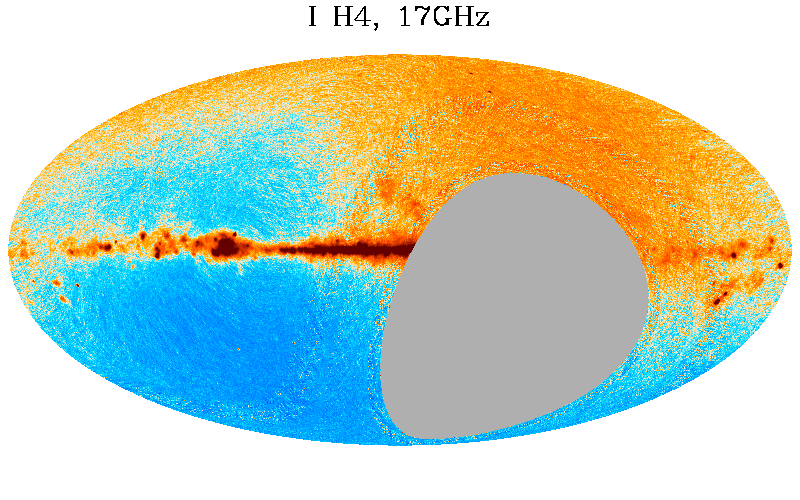} \includegraphics[width=0.24\textwidth]{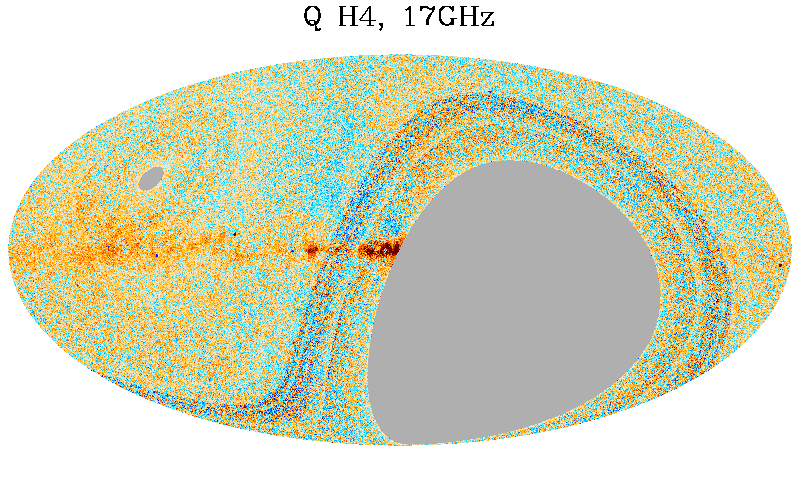} \includegraphics[width=0.24\textwidth]{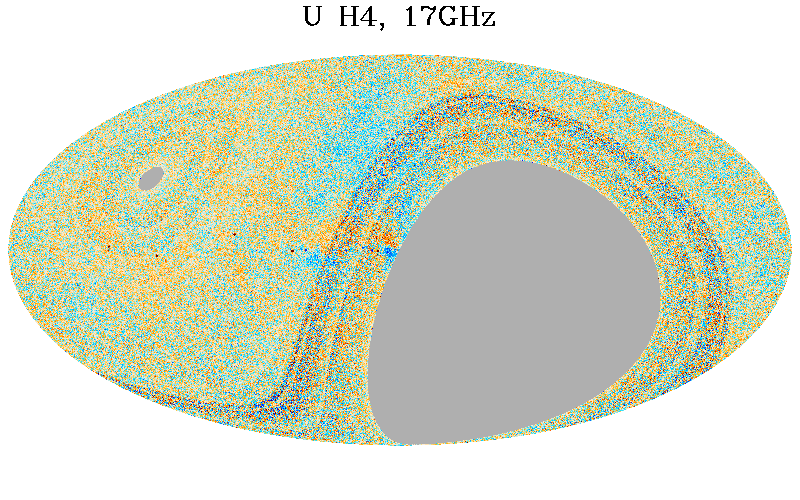}
    \includegraphics[width=0.24\textwidth]{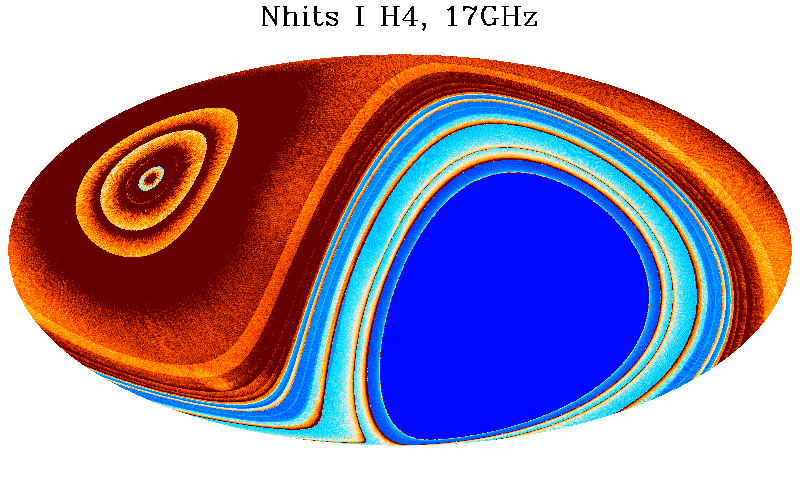}
    
    \includegraphics[width=0.24\textwidth]{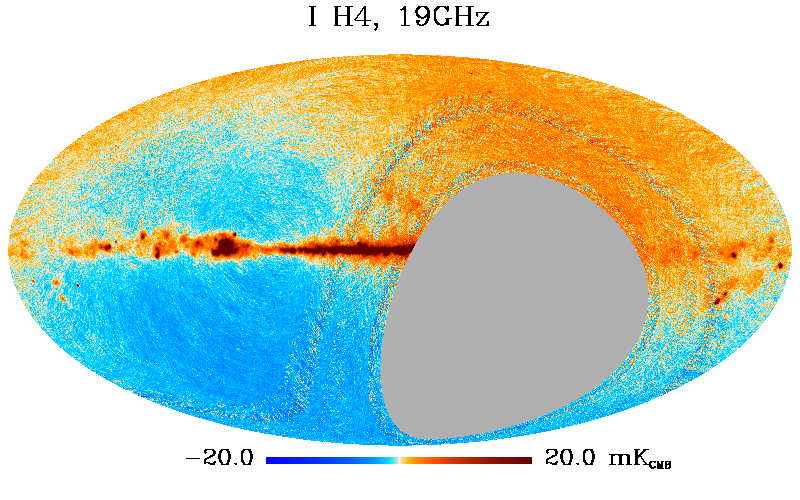} \includegraphics[width=0.24\textwidth]{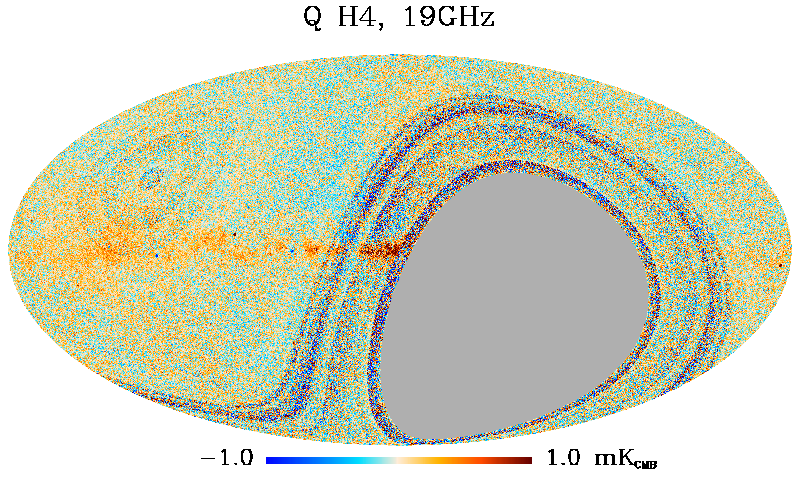} \includegraphics[width=0.24\textwidth]{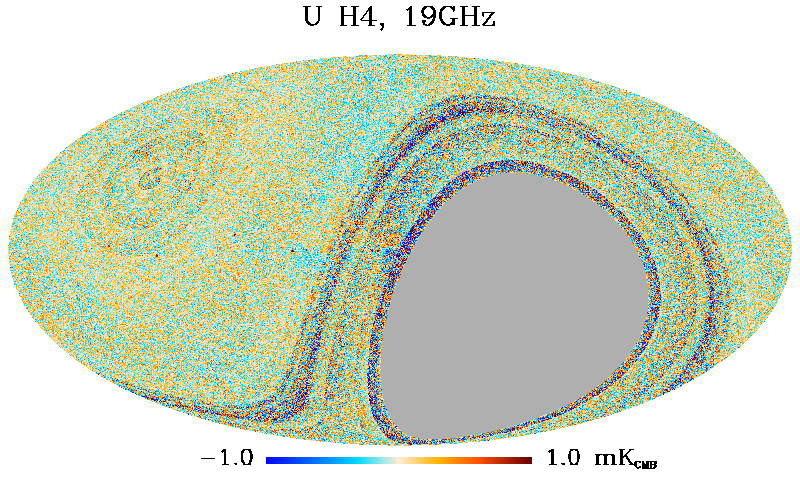}
    \includegraphics[width=0.24\textwidth]{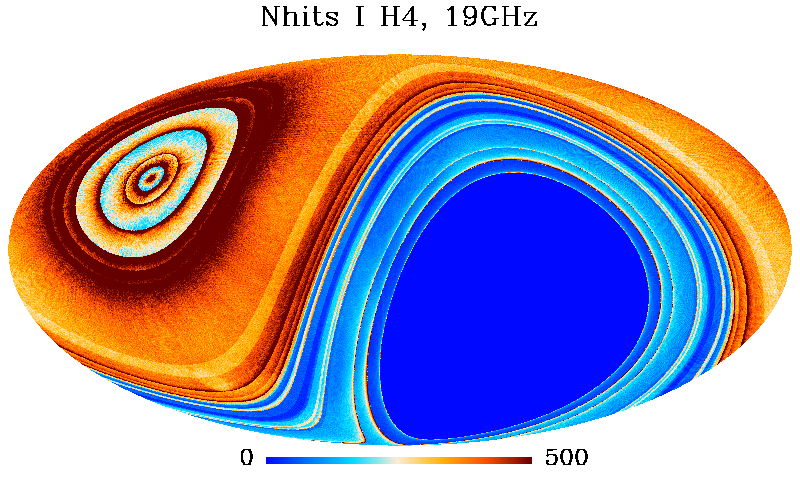}

\caption{Reconstructed \textit{I}-\textit{Q}-\textit{U} maps of the simulated microwave sky, in galactic coordinates, as they would be observed by QUIJOTE-MFI across the full sky area accessible from the Teide observatory, after processing the corresponding TOD with \namecode{}. The rows show, in order, the maps at 11, 13, 17 and 19\,GHz, of horn number 3 for the low frequencies, and of horn number 4 for the high frequencies. The columns show, in order, the maps of \textit{I}, \textit{Q}, \textit{U},  and the number of hits of the intensity, at the correspondent frequency. For display purposes, the maps are degraded to $N_{\rm side}=256$ (pixel size    $\sim 13.7$\,arcmin).}
\label{fig_maps}
\end{figure*}

The main result of this paper is the validation and characterization of the \namecode{} map-making code, at the map and angular power spectrum level, by using realistic simulations of the QUIJOTE-MFI wide-survey described in Sec.~\ref{sec_maps}.
Fig.~\ref{fig_maps} shows the reconstructed \textit{I}, \textit{Q} and \textit{U} maps obtained with \namecode{} at the four MFI frequencies, for the case of realistic noise levels (including both white noise and \oof noise).

For definiteness, we focus our discussion on the two extreme frequencies of the MFI: at 11\,GHz, where the sky signal is brighter, and at 19\,GHz, where the sky signal is fainter. 
A detailed study of the 11\,GHz map, including a comparison with the input data and the residual levels of the reconstruction, is shown in Fig.~\ref{fig_maps_h311}.
Sec.~\ref{sec_maps} contains a characterization of these residual maps, including real space statistics. The angular power spectra of the 11 and 19\,GHz maps, together with an analysis of the signal error within the mask in Fig.~\ref{fig_mask}, are shown in Sec.~\ref{sec_cl} (Fig.~\ref{fig_cl} and \ref{fig_signal_error}), while a detailed characterization of the transfer function at 11\,GHz is shown in Sec.~\ref{sec_transfer_function} (Fig.~ \ref{fig_transfer_function_nonoise}). Afterwards, in Sec.~\ref{sec_fit_results} we present the validation of the fitting of a template function implemented in \namecode{} (as described Sec.~\ref{sec_fitfunc} and \ref{sec_template}). Finally, we tested with simulations the detection of the CMB anisotropies, as it is done with the real wide-survey data. This is shown in Sec.~\ref{sec_cmb} (Fig.~\ref{CMB_mask} and \ref{fig:CMB_crosscorr}).

\subsection{Reconstructed maps and real space statistics}
\label{sec_maps}

\begin{figure*}
    \centering
    \includegraphics[width=0.3\textwidth]{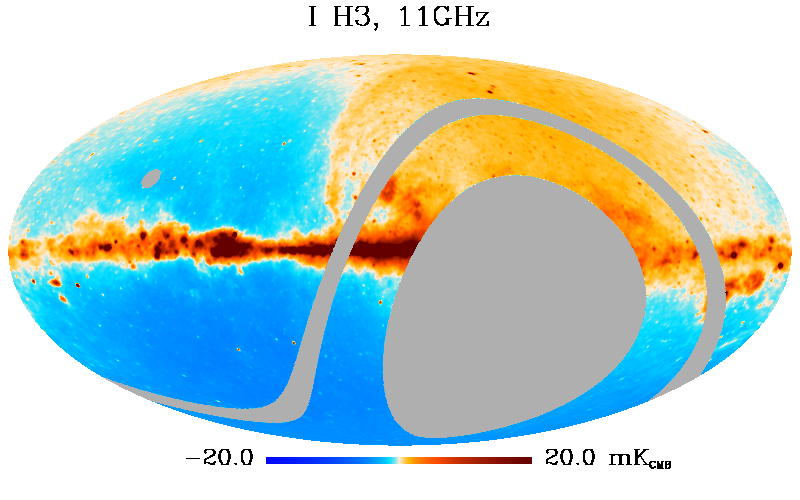} \includegraphics[width=0.3\textwidth]{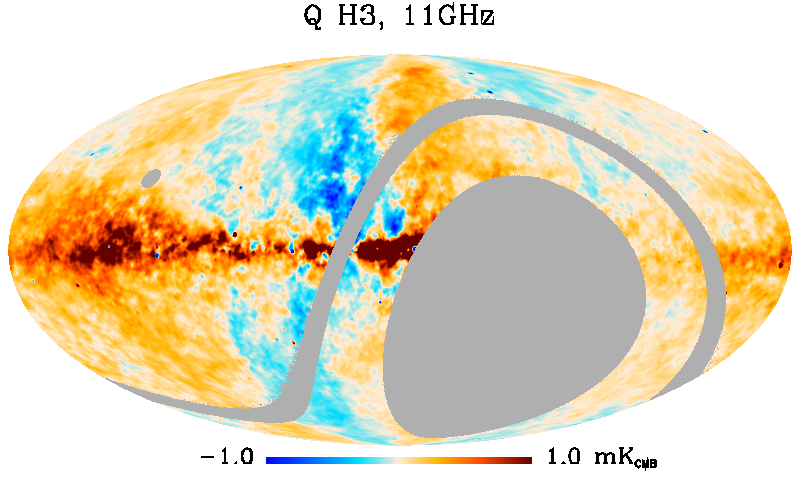} \includegraphics[width=0.3\textwidth]{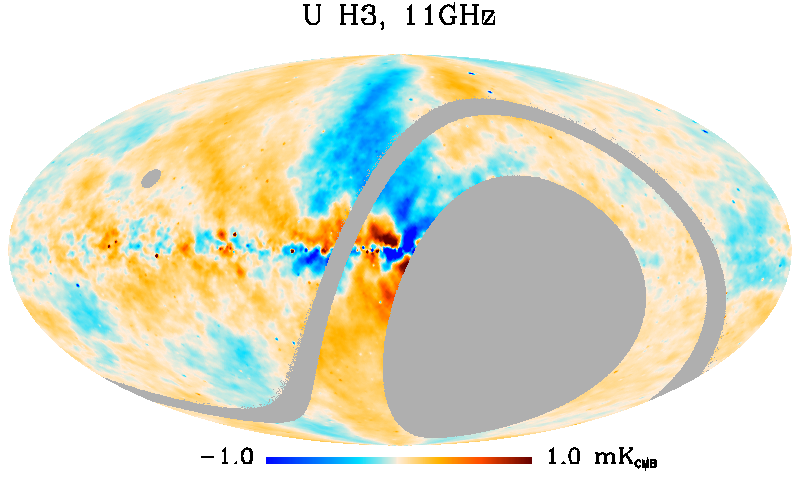}

    \includegraphics[width=0.3\textwidth]{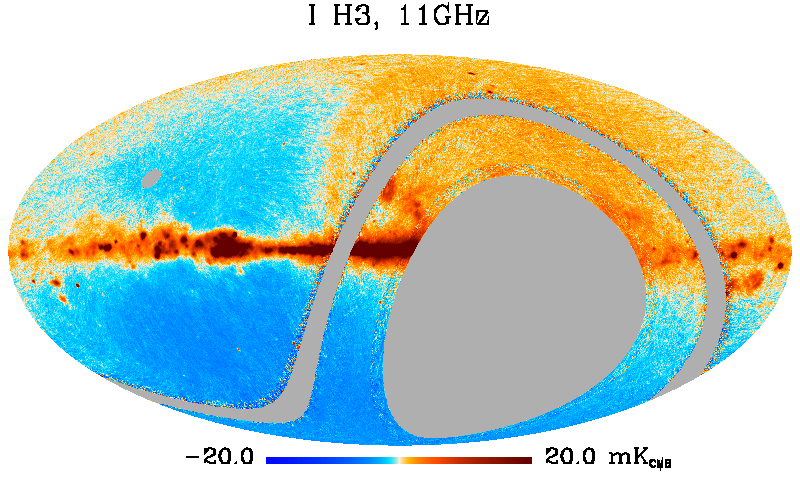} \includegraphics[width=0.3\textwidth]{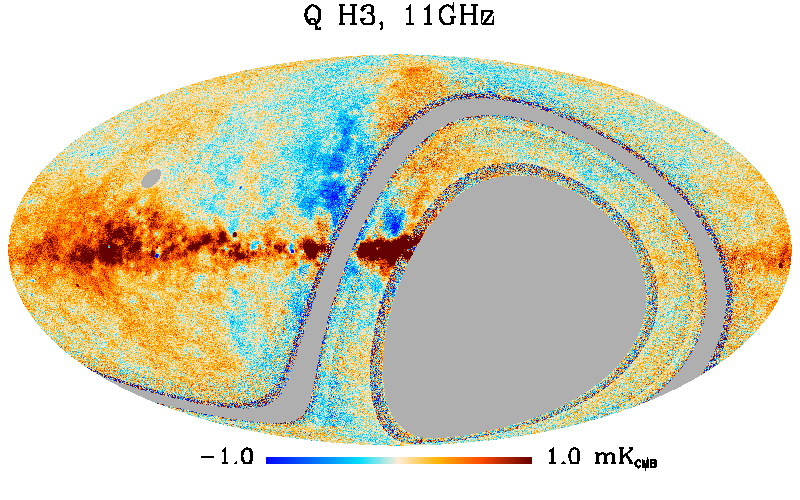} \includegraphics[width=0.3\textwidth]{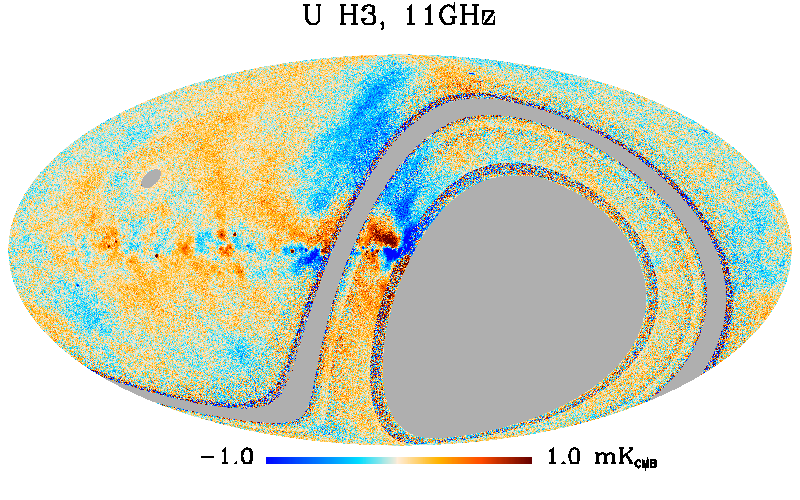}
    
    \includegraphics[width=0.3\textwidth]{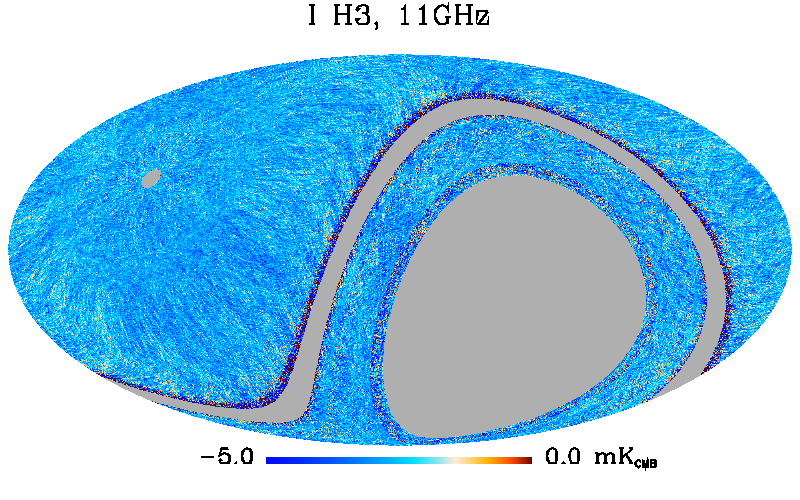} \includegraphics[width=0.3\textwidth]{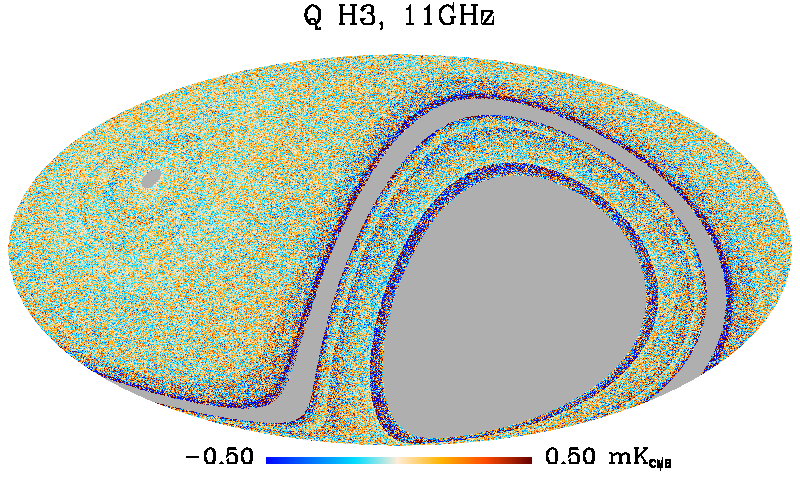} \includegraphics[width=0.3\textwidth]{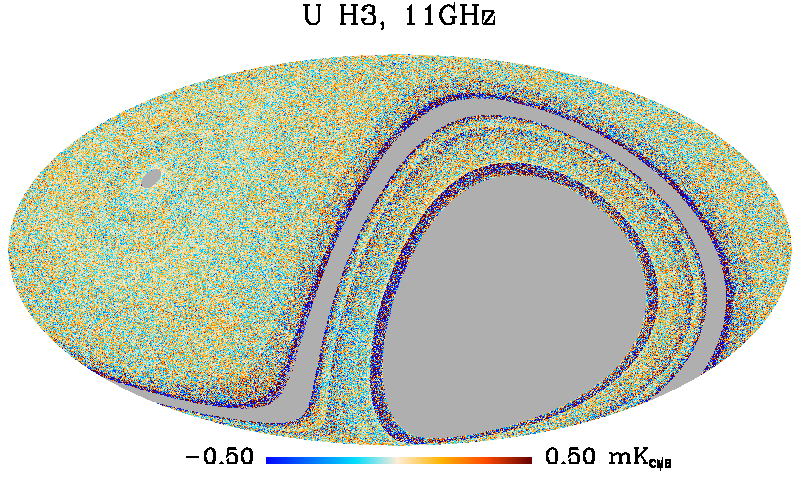}
    
    \caption{Stokes \textit{I}, \textit{Q}, and \textit{U} sky maps at 11\,GHz, from horn number 3.  We show in the top row the input sky with CMB dipole included, in the central row the reconstructed maps obtained with \namecode{} from the TODs containing realistic white plus \oof noise, and in the bottom row the residual noise maps, obtained as the difference between the two maps above. For display purposes, the maps are degraded to $N_{\rm side}=256$.}
    \label{fig_maps_h311}
\end{figure*}

\begin{figure*}
    \centering
    \includegraphics[width=0.49\textwidth]{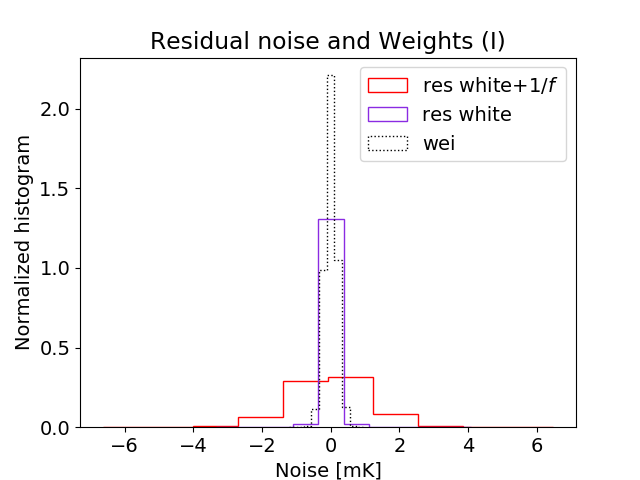}
    \includegraphics[width=0.49\textwidth]{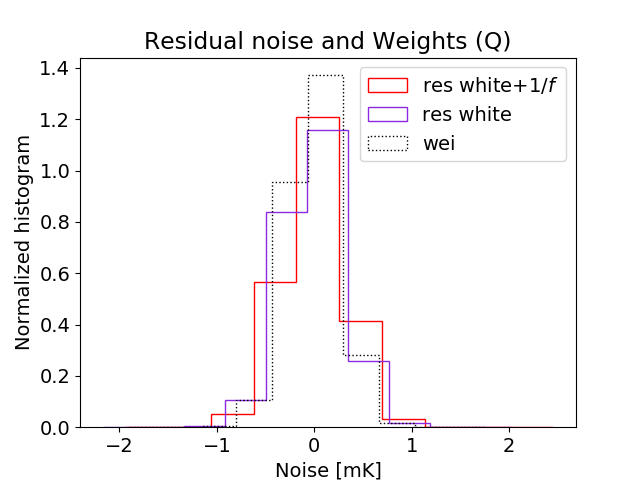}
    \caption{Histogram of the noise distribution in intensity (\textit{I}) and polarization (\textit{Q}), obtained with the residual maps, and compared with the white noise expectation obtained from the weight maps (black dotted line). We show in purple the results obtained with simulation at 11\,GHz containing white noise only, and in red the case where $1/f$ noise is also present (corresponding to the maps shown in Fig.~\ref{fig_maps_h311}).  }
    \label{fig:histo}
\end{figure*}

We have tested our code using the realistic simulations of the QUIJOTE-MFI wide-survey described in Sec.~\ref{Sec_simu}, which  include sky signal, CMB dipole, CMB anisotropies and noise. We consider three set of simulations: one with no-noise, one with white noise only, and the third one with realistic \oof noise (hereafter, no-noise, white and white plus $1/f$, respectively, see details in Sec.~\ref{noisesim}). 
As described in Sec.~\ref{sec:binningIQU}, from the four channels $(\textbf{V}_{1},\,\textbf{V}_{2},\,\textbf{V}_{3},\,\textbf{V}_{4})$ of a determined horn and frequency of the MFI, \namecode{} produces two maps: $m_1$ for the pair of channels $(\textbf{V}_{1},\,\textbf{V}_{2})$, and $m_2$ for  $(\textbf{V}_{3},\,\textbf{V}_{4})$. We call $w_1$ and $w_2$ the correspondent weight maps, which are obtained as the inverse of the variance maps defined in Sec.~\ref{sec_map} (Eq.~\ref{Ierr} and \ref{QUerr}). The two maps and weights are constructed with two independent runs of the code, and are combined a posteriori with a weighted average:
\begin{equation}
    m = \frac{m_1 w_1 + m_2 w_2}{ w_1+ w_2}
\label{eq:wei_avg_map}
\end{equation}
The weight map of the combination is given by:
\begin{equation}
    w =  w_1+ w_2
\label{eq:wei_avg_wei}
\end{equation}
In summary, from each horn of the MFI, we obtain two combined frequency maps, one for each of the two frequencies of a selected horn. For example, from horn number 3 we construct one combined map at 11\,GHz and one at 13\,GHz, and from horn number 4 we obtain one combined map at 17\,GHz and one at 19\,GHz. 

We construct the maps with \namecode{} using $N_{\rm side}=512$, in order to have an appropriate sampling of the MFI beams, and a baseline length of 2.5\,s \citep{alba}. 
The maps of \textit{I}, \textit{Q}, and \textit{U}, and the number of hits ($n_{\rm hit}$) of 40\,ms time samples in pixels of $N_{\rm side}=512$, are shown in Fig.~\ref{fig_maps}, for the case of white plus \oof noise. We show the maps obtained from horn number 3 at  11 and 13\,GHz, and from horn number 4 at 17 and 19\,GHz. In this figure, we can see how simulated \textit{I}-\textit{Q}-\textit{U} maps of the microwave sky, in galactic coordinates, would be observed by QUIJOTE-MFI at 11, 13, 17 and 19\,GHz, across the full sky area accessible from the Teide observatory, after processing the corresponding TOD with \namecode{}. 

A quick look at the maps shows the bright emission of the CMB dipole in intensity. As expected, the CMB dipole has the same amplitude at each frequency, contrarily to the synchrotron emission of the Galaxy, which decreases with frequency. This reconstruction of the CMB dipole is encouraging, as indicates that \namecode{} reconstructs with good precision the sky signal even at large angular scales. 
We recover 100\,\% of the injected CMB orbital plus solar dipole, with a precision of the order of $0.1$\,\%, at all frequencies, and independently on the noise cases (either no-noise, white, or white plus $1/f$ noise).

The grey regions in the maps correspond to the sky area that is not observed by QUIJOTE, including a circle around the North Celestial Pole (NCP), the southern sky at low declination, and an intermediate band close to Dec.$=0$\,deg, that must be flagged due to interference by geostationary satellites at QUIJOTE frequencies, especially at 11 and 13\,GHz (see details in Rubi\~no-Martin et al. in prep).

Some of the maps show evident ring structures, located around the NCP, and at low declination, in the band that crosses the Galactic center. The rings correspond to discontinuities in the sky coverage (see $n_{\rm hit}$ map in the last column of Fig.~\ref{fig_maps}), and, consequently, to variations of the noise properties across the map. This is a direct consequence of the scanning strategy. Indeed, the rings delimit the declination bands that are observed at different elevations. The very low declination regions can only be accessed with low elevation scans, and therefore the amount of data there is much smaller than in the center of the map, and the noise is larger. Also the high declinations are only accessible by low elevation scans, however in this case there is a projection effect that compensates for that, producing more hits approaching the North Celestial Pole.

For definiteness, we present a more detailed analysis with the maps at one selected frequency, at 11\,GHz, where we have the best signal-to-noise.  Fig.~\ref{fig_maps_h311}  allows a visual comparison to be made between the input sky signal of the simulations and the maps constructed with \namecode{} (shown in Fig.~\ref{fig_maps}). In the upper line we show the \textit{I}, \textit{Q} and \textit{U} input sky maps with the CMB dipole included, in the central line the maps reconstructed with \namecode{} from simulations with white plus \oof noise, and in the bottom line the difference between the recovered map and the input sky, being them residual noise maps. We can notice from the difference maps that there are not evident sky signal residuals, either at small or large angular scales. This demonstrates that \namecode{} reconstructs correctly the injected sky signal. A more quantitative analysis at the angular power spectrum level is presented in Sec.~\ref{sec_cl} and \ref{sec_transfer_function}.

However, we note that the residual noise maps of the intensity present \oof correlated noise structures that \namecode{} was not able to cancel perfectly. On the other hand, the residual noise in polarization is apparently consistent with white noise. This can be quantified using the histograms shown in Fig.~\ref{fig:histo}, where we compare the actual distribution of the noise obtained from the residual maps, with the expected white noise level. The latest can be computed directly from the weight maps (or estimated from a Gaussian realization with variance given by the inverse of the weight map), and for this reason, it is labelled as "wei" in the figure. 
In the histograms we can see that the distribution of the noise in intensity (upper panel, red thick line) is wider than the expected white noise levels given by the weights (upper panel, black dotted line, while the noise in polarization (lower panel, red thick line) is statistically consistent with the expected white noise reference (lower panel, black dotted line). 
In other words, in polarization, \namecode{} is performing as   
optimal, in the sense that it recovers the expected white noise level. However, in intensity, there is a measurable correlated noise residual in the map-making solution, as expected for the QUIJOTE-MFI data. 

This result can be easily explained in terms of the actual correlated noise injected in the TODs as compared with the baseline length that we used. We estimated the baselines for both the intensity and polarization data using a length $t_{\rm b}=2.5$\,s, which corresponds to a frequency of the baselines $f_{\rm b}=1/2.5$\,Hz\,$=0.4$\,Hz. The knee frequency of the injected noise in intensity is $f_{\rm k}=20$\,Hz, which is much higher than the frequency of the baselines. In the polarization data, instead, we have a \oof noise component with a knee frequency $f_{\rm k}=0.3$\,Hz, which is comparable but lower than $f_{\rm b}$. 
A shorter baseline length could be an option to improve the noise cleaning of the intensity maps, but short baselines also imply a poorer reconstruction of the large angular scales, in particular if the prior do not perfectly match the actual noise, (see \cite{Kurki-Suonio} for a study with ${\bf C}_{\rm a}^{-1}=0$). We tested, for example, a baseline length $t_{\rm b}=1$\,s, and we obtained a more noisy reconstruction of the large angular scales signal, as compared with $t_{\rm b}=2.5$\,s. This is in agreement with \cite{alba}, where they estimate the optimal baseline length to be $t_{\rm b}=2.5$\,s. Moreover, the scan speed and the beam size set a lower limit for the baseline length, which is given by some multiple of $t_{\rm B} = \mathrm{FWHM} / (v  \cos(el))$, the time that the telescope takes to scan one beam FWHM, with azimuthal scan speed $v$, and at the constant elevation $el$.
Typical values for QUIJOTE are $v=12$\,deg/s, $el=60$\,deg and FWHM=1\,deg, which gives $t_{\rm B}=0.17$\,s. 
This means that, in order to preserve structures with sizes of few beams (e.g., 5--10 FWHMs) and not to confuse them with features associated to noise in the maps, the baseline length should be longer than $5$--$10 t_{\rm B}$, i.e. 0.8--1.7\,s. This highlights the importance of scanning the sky as fast as possible, in order to be able to suppress efficiently the \oof noise.


\subsection{Validation with angular power spectra} 
\label{sec_cl}

\begin{figure}
    \centering
    \includegraphics[width=0.4\textwidth]{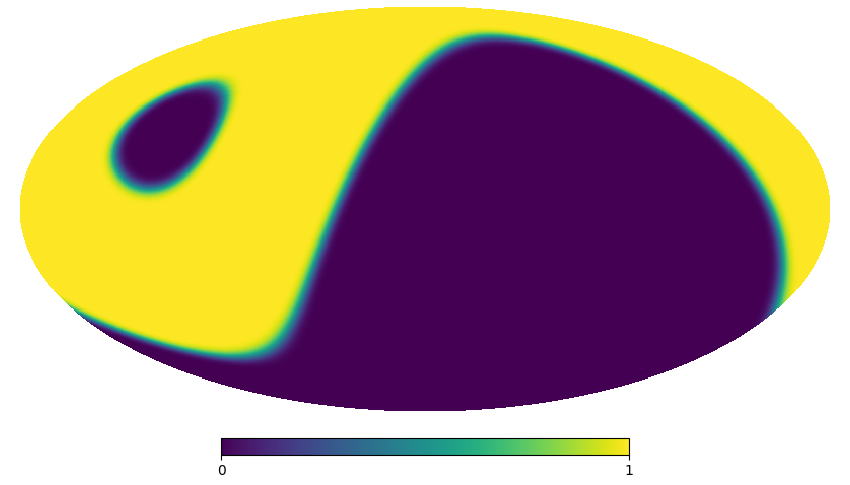}
    \caption{Mask adopted for angular power spectra estimations.}
    \label{fig_mask}
\end{figure}
\begin{figure*}
    \centering
    \includegraphics[width=0.49\textwidth]{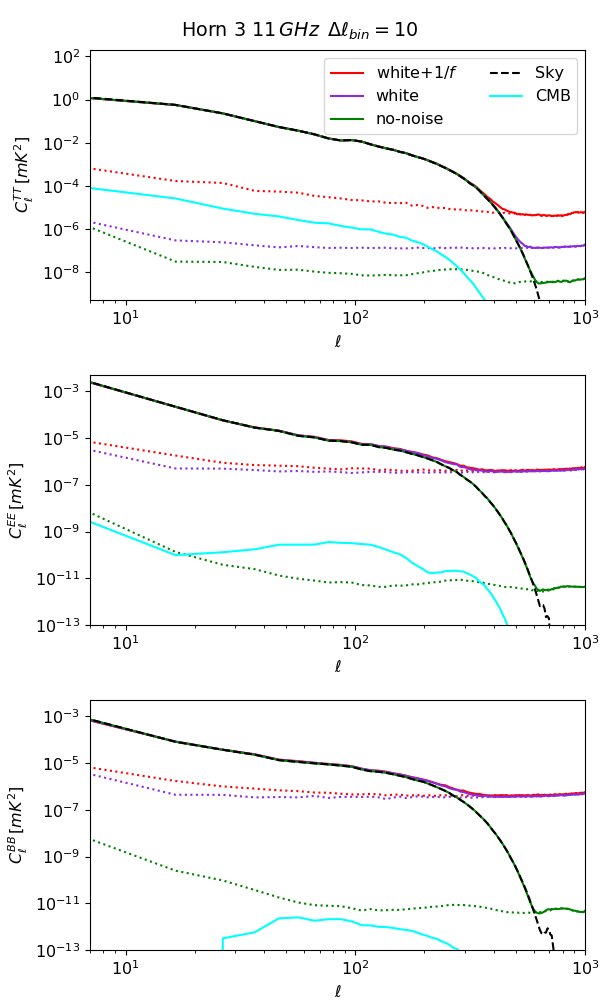}
    \includegraphics[width=0.49\textwidth]{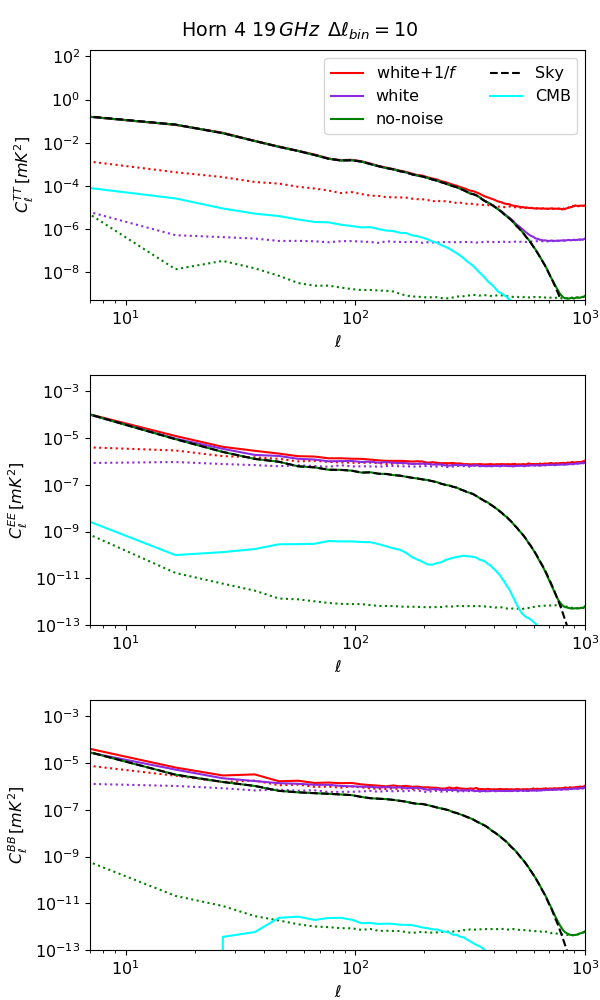}
    \caption{
    Angular power spectra of the two selected frequency maps, 11\,GHz (left) and 19\,GHz (right), from horns 3 and 4 respectively. Different colors correspond to the maps simulated with different noise properties: in red the case with white plus \oof noise, in purple the simulation with white noise only, and in green the simulation without noise. The light blue line shows the power spectrum of the CMB anisotropies, convolved with the beam window function of the MFI at the correspondent frequency. We show, for the two frequencies, the TT, EE and BB auto power spectra of the maps,  respectively in the top, central and bottom position. The different lines represent the $C_{\ell}$ of the recovered simulated map (thick lines), of the input sky map (dashed black line) and of the residual noise map (map minus sky; dotted lines).  The \cl{}'s are not corrected by the beam window function, while the pixel window function correction is applied.  }
    \label{fig_cl}
\end{figure*}

\begin{figure}
    \centering
    \includegraphics[width=0.49\textwidth]{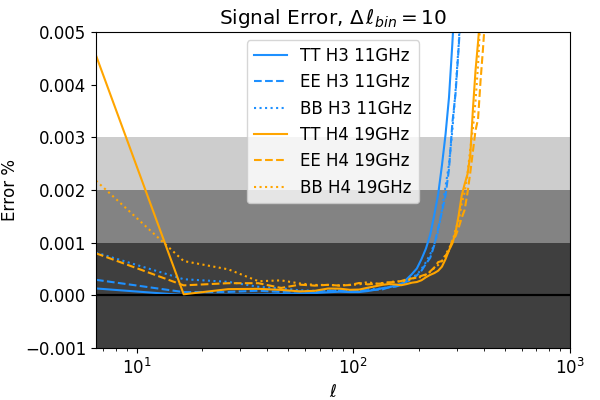}
    \caption{Percentage signal error of TT (thick line), EE (dashed line), and BB (dotted line), at  $11$\,GHz (blue) and $19$\,GHz (orange).  It is computed as the percentage ratio between the angular power spectrum   of the signal error map and of  the recovered map, in the case with no-noise. The shaded areas mark the deviation of the signal error from zero by $(0.001,\,0.002,\,0.003)\%\,$. }
    \label{fig_signal_error}
\end{figure}

We analyze here the angular power spectra of the simulated QUIJOTE-MFI wide-survey. The $C_{\ell}$'s are computed with the publicly available code \xpol,\footnote{\href{https://gitlab.in2p3.fr/tristram/Xpol}{https://gitlab.in2p3.fr/tristram/Xpol}} which is based on a pseudo-$C_{\ell}$ estimator, and accounts for incomplete sky coverage (\citealt{Tristam2005}). 

Pseudo-$C_{\ell}$ is a very useful tool for computing the angular power spectra of maps with incomplete sky coverage, and with a large number of pixels. However, it is potentially affected by residual mode coupling at low multipoles, particularly if the map contains a dipole. Therefore, we need to carefully remove the CMB dipole from our simulations before computing the power spectrum. For the analysis in intensity, we fit and remove a dipole component from the simulated map subtracted from the foregrounds and the CMB anisotropies, and we use the dipole subtracted residual to characterize the noise.

For this work, we use a mask of the high signal-to-noise QUIJOTE sky area, that encompasses the declination range Dec.$ \in [5^{\circ},70^{\circ}]$, as shown in Fig.~\ref{fig_mask}. We applied to the mask a five degrees apodization with a cosine function, with the apodization routine of the {\sc NaMaster}\footnote{\url{https://github.com/LSSTDESC/NaMaster}} publicly available code (\citealt{NaMaster}).

We show in Fig.~\ref{fig_cl} the angular power spectra of two selected frequency maps, those with the highest and the lowest signal-to-noise:  11\,GHz of horn 3 and 19\,GHz of horn 4. The plots show, for the two frequencies (11\,GHz on the left and 19\,GHz on the right), the TT, EE and BB auto power spectra of the maps (respectively in the top, central and bottom position), where EE and BB represent the auto-spectra of the commonly called polarization E and B-modes (\citealp{kamionkowski1997a}). The different lines in these plots represent the $C_{\ell}$ of the map recovered with \namecode{} (thick lines), of the input sky map (dashed black line) and of the residual noise map (dotted lines). Finally, the colors represent the noise properties of different simulations: in red we represent the simulation with white plus \oof noise, in purple  with only white noise, and in green the result from a simulation without noise.
For comparison purposes, the light blue line depicts the power spectrum of the simulated CMB map, convolved with the beam window of the MFI at 11 and 19\,GHz, and computed within the same sky mask (Fig.~\ref{fig_mask}). In this particular simulation, the CMB BB signal is entirely due to lensing (no tensor modes). The \cl{}'s are not corrected by the beam window function, while the pixel window function correction is applied. 

Let us now describe the results represented in the power spectra of Fig.~\ref{fig_cl}. First, we analyze the angular power spectrum of the signal error, which is the map of the residual (map minus input sky) when no noise is added into the simulated data-set (\citealt{Planck_LFI_30}). The signal error is represented by the green dotted lines in Fig.~\ref{fig_cl}, and it quantifies the error introduced by the map-making 
when reconstructing the sky signal, despite the absence of noise. We can notice that the TT, EE and BB signal error of \namecode{} lies several orders of magnitude below the angular power spectrum of the map (green thick lines) at the relevant multipoles for QUIJOTE ($10<\ell<400$), meaning that the error made by \namecode{} when reconstructing the sky signal into the map is low with respect to the level of the signal itself. We show in Fig.~\ref{fig_signal_error} the TT, EE and BB percentage signal error angular power spectra. We can see here that the signal error is very small, being lower than $0.005\%$ at  multipoles $\ell < 400$, for TT, EE and BB, while in the range $20< \ell < 200$ the signal error is lower than $0.001\%$. It increases then in the two extreme regimes: at very low multipoles, showing that the reconstruction of the large angular scales is well under control down to $\ell = 10$, and at high multipoles $\ell > 400$,  where the signal drops due to the effect of the window function.

We can now compare the simulations with no noise (green lines in Fig.~\ref{fig_cl}) with the simulations containing white noise (purple lines in Fig.~\ref{fig_cl}).  The angular power spectrum of the residual in the white noise only case is approximately flat at multipoles  $\ell > 20$, as expected for white noise. In the low multipole range, instead, it shows a mild growth, which is probably related to the increase of the signal error at $\ell < 20$, particularly in the intensity case, although some residual 
mode-coupling due to the finite sky coverage could be present.

Finally, we compare with realistic simulations containing white plus \oof noise, which are represented by the red color lines in Fig.~\ref{fig_cl}. In intensity (TT), we can notice the effect of the \oof contamination in two different multipoles regimes. First, at low multipoles, we can clearly see the typical $1/\ell$-like rise of the power spectrum, which is due to the residual \oof{}correlated noise structures in the intensity maps. Second, at high multipoles, we can observe that the noise level of the \oof simulation is about two orders of magnitude higher than that of the white noise only simulation, while we could expect them to be comparable. 
This effect is due to the high \oof knee frequency of the intensity data, as compared with the sampling frequency of the TOD. Indeed, since we simulate \oof noise with a knee frequency of $f_{\rm k}=20$\,Hz, and we bin the TOD in $40$\,ms time intervals, we have \oof noise drifts even within one single time bin. This artificially enhances the rms of the data samples ($\sigma_i$ of Eq.\,\ref{weights}) and therefore also the white noise level of the intensity maps. 

In polarization (EE and BB), where the simulated \oof noise is small, we can see that the white plus \oof noise power spectrum  is close to the white noise only case, indeed overlapping at high multipoles. This means that the selection of the baseline length is producing (nearly) optimal results, in the sense that the code recovers the white noise levels at high multipoles. 
However, also in this case, we observe a rise of the noise angular power spectrum at low multipoles, which is due to a combination of residual large-scales correlated noise and signal error.

To conclude, we can notice that the power spectrum of the sky signal is well reconstructed, if the signal to noise is sufficiently good. In the realistic case with \oof noise, at 11\,GHz (red line in the left panels of Fig.~\ref{fig_cl}), the TT angular power spectrum of the map is above the noise contribution
up to $\ell \sim 300$, while the EE and BB  \cl's are well reconstructed up to $\ell \sim 100$. At $19$\,GHz, where the sky signal is weaker, the quality of the reconstruction is worse. At this frequency we have a good reconstruction of TT up to $\ell \sim 100$, while the EE and BB power spectra are noise dominated at all multipoles.

\subsubsection{Parametric fit of the noise angular power spectrum} \label{fit_nl}
We fit the noise angular power spectra with the following empirical model:
\begin{equation}
    C_{\ell}= C_{\rm w}\left(1+\left(\frac{\ell_{\rm k}}{\ell}\right)^{\alpha}\right),
    \label{eq:fit_cl}
\end{equation}
in analogy to the \oof noise in the frequency space given by Eq.\,\ref{ps_teo_whiteplusoof}. The parameter $C_{\rm w}$ represents the white noise level of the maps. In practise, it 
can be obtained as the average of the angular power spectrum at high multipoles ($\ell \in [700,800]$ for TT and $\ell \in [400,500]$ for EE and BB). Note that $C_{\rm w}$ can be translated into the commonly used quantity $\sigma_{\text{1-deg}}$, which is the rms of the map in a 1-degree beam (see numerical values in Tab.\,\ref{tab:fit_params_cl}), with the relation $\sigma_{\text{1-deg}}=\sqrt{C_{\rm w}/\Omega_{\text{1-deg}}}$, where $\Omega_{\text{1-deg}}$ is the solid angle of a Gaussian beam with a FWHM of 1-degree. The parameter $\ell_{\rm k}$ is the knee-multipole between a $1/\ell$ and a flat (white) regime. The knee-multipole $\ell_{\rm k}$ is obtained analytically after fitting a linear slope in $\log_{10}(C_{\ell}-C_{\rm w})\,vs\, \log_{10}(\ell)$, in a range of intermediate multipoles $\ell \in [10,100]$ for TT, and $\ell \in [10,80]$ for EE and BB. Being $\alpha$ the angular coefficient of the linear slope mentioned above, and \textit{q} the fitted intercept,\footnote{Analytically, the intercept is given by $q=\log_{10}(C_{\ell}(1)-C_{\rm w})=\log_{10}(C_{\rm w})+\alpha \log_{10}(\ell_{\rm k})$.} the $\ell_{\rm k}$ is given by:
\begin{equation}
    \ell_{\rm k} = 10^{\left(\frac{q-\log_{10}(C_{\rm w})}{\alpha}\right)}.
\end{equation}
Table~\ref{tab:fit_params_cl} reports the $C_{\rm w},\, \ell_{\rm k}$ and $\alpha$ parameters extracted from the simulations with white plus \oof{}noise, for TT and EE. As expected, the \oof{}noise in intensity is reflected into a large  $\ell_{\rm k}$, while in polarization, where the \oof{}noise is low, also the $\ell_{\rm k}$ is much lower. It is interesting to notice that the noise parameters $\gamma = 1.5$ and $f_{\rm k}=20$\,Hz injected in the intensity simulation (see Tab.\,\ref{Table_noise_sim}) are translated into a $\alpha \approx 1.2$ and $\ell_{\rm k} \approx 400$ in the angular power spectrum domain. Analogously, the parameters $\gamma=1.8$ and $f_{\rm k}=0.3$\,Hz used for the simulated noise in polarization correspond, at the angular power spectrum level, to  $\alpha \approx 1.3$ and $\ell_{\rm k} \approx 40$.

\begin{table}
    \centering  
\begin{tabular}{ccccc}
\hline\hline
Horn & Frequency [GHz] & $\sigma_{\text{1-deg}}\,[\mu K]$ & $\alpha$ & $\ell_{\rm k}$\\ 
\hline
\hline
\multicolumn{5}{c}{TT} \\
\hline
\hline
3 & 11.0 & 98.5 & 1.31 & 370.3  \\
3 & 13.0 & 87.7 & 1.22 & 390.2  \\
\hline
4 & 17.0 & 128.5 & 1.19 & 428.7 \\
4 & 19.0 & 142.4 & 1.41 & 323.5 \\
\hline
\hline
\multicolumn{5}{c}{EE} \\
\hline
\hline
3 & 11.0 & 33.0 & 1.45 & 37.1\\
3 & 13.0 & 29.3 & 1.26 & 39.2\\\hline
4 & 17.0 & 44.6 & 1.17 & 38.4\\
4 & 19.0 & 45.0 & 1.31 & 38.1\\
\hline\hline
\end{tabular}
\caption{Fitting of the TT and EE noise angular power spectra of the simulation with white plus \oof noise, according to Eq.~\ref{eq:fit_cl}.}\label{tab:fit_params_cl}
\end{table}

\subsection{Transfer function} \label{sec_transfer_function}

\begin{figure}
\includegraphics[width=0.49\textwidth]{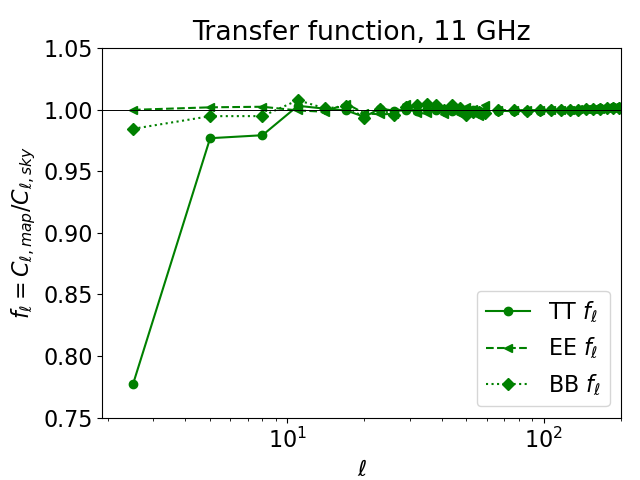} 
\caption{Transfer function of the \namecode{}  map-making code, obtained from a simulation of the QUIJOTE-MFI wide survey, at 11\,GHz, in absence of noise  and CMB dipole. The thick line represents the intensity (TT), and the dashed and dotted lines, respresent the polarization (respectively EE and BB).}
\label{fig_transfer_function_nonoise}
\end{figure}

We quantified the large angular scale suppression introduced by \namecode{} and by the wide-survey scanning strategy, by performing a study of the transfer function of the simulated wide-survey maps (Fig.~\ref{fig_transfer_function_nonoise}). We used the simulations in the ideal case with no-noise, for horn number 3 at 11\,GHz, as presented in Sec.~\ref{sec_cl}, with no CMB dipole included. 

The transfer function is computed as the ratio between the $C_{\ell}$ of the reconstructed map and the $C_{\ell}$ of the input sky, both computed within the mask in Fig.\,\ref{fig_mask}. In order to control the possible residual mode coupling at large angular scales that could affect the pseudo-$C_{\ell}$ estimator, we computed the (binned) low multipoles points of the angular power spectra ($\ell < 60$) with a fast and robust implementation of a quadratic maximum likelihood $C_{\ell}$ estimator (\texttt{ECLIPSE}, \citealp{Bilbao2021}), after degrading the maps to $N_{\rm side}=32$ (pixel size $\sim$1.8\,deg). The power spectra at multipoles higher than $\ell = 60$ are computed with the pseudo-$C_{\ell}$ code \xpol.

The results are shown in Figure~\ref{fig_transfer_function_nonoise}, for TT, EE and BB. We can observe that in TT \namecode{} recovers $\sim 100$\% of the Galactic signal at multipoles $\ell>10$, while there is a loss of power of $\sim 3$\% at multipoles $2 < \ell < 10$, and of $\sim20$\% at $\ell=2$. However, as it can be clearly seen in the intensity residuals shown in Fig.~\ref{fig_maps_h311} that, when we include the CMB dipole ($\ell = 1$), it is perfectly recovered at the map level. We understand that \namecode{} is able to reconstruct the CMB dipole because its signal is sufficiently high, while the Galactic signal at angular scales with $2 < \ell < 10$ has less power, and is therefore more complex to reconstruct.\footnote{Note that for the computation of the transfer function we use a map that does not include the CMB dipole, in order to avoid any possible mixing or the large angular scale modes.}  In polarization, the transfer function of EE and BB shows that with \namecode{} we can recover $\sim 100$\% of the signal at $\ell>8$, while the loss of power at lower multipoles is not larger than $2$\%. This behaviour of \namecode{} in polarization is particularly promising in prospective for future works aimed to detect the primordial B-modes. Similar results are found for the other simulated MFI frequencies.

\subsection{Fit of a template function}\label{sec_fit_results}

\begin{table}
\centering 
\resizebox{0.48\textwidth}{!}{%
\begin{tabular}{ccccc|cc}\hline\hline
Noise & Fit & CMB dip.  & $\sigma^{\rm MC}_A$ & $\sigma^{\mathrm{analytic}}_{A}$ & $<A>$ & $\sigma_{<A>}$ \\ \hline\hline
\multirow{3}{3em}{White} &
Atmos. & No &  0.0009 & 0.0009 & 1.0000 & 0.0001 \\ 
& Atmos. & Yes  &  0.0009 & 0.0009 & 1.0038 & 0.0001 \\ 
& CMB dip. & Yes &  0.011\phantom{0} & 0.012\phantom{0} &  0.992\phantom{0} & 0.002\phantom{0} \\ \hline
\multirow{3}{3em}{$1/f$} &
Atmos. & No &  0.0233 & 0.0009 & 1.0018 & 0.0033 \\ 
& Atmos. & Yes &   0.0245 & 0.0009 & 1.0057 & 0.0035 \\ 
& CMB dip. & Yes & 0.28\phantom{0}\phantom{0} & 0.01\phantom{0}\phantom{0} & 1.00\phantom{0}\phantom{0} & 0.04\phantom{0}\phantom{0} \\ \hline\hline
\end{tabular}}
\caption{Result of the fitting technique during map-making, for a stable plane-parallel atmosphere and for the CMB dipole, for two noise cases: white noise only and white+$\oof$ noise. Realistic simulations have been used to estimate the uncertainty for the fitted amplitudes $A$, by generating 50 independent TOD noise realizations, added on top of the same simulated TOD containing the sky signal.  This table reports  the uncertainty for a single estimate of $A$ obtained as the dispersion of the MC realizations ($\sigma^{\rm MC}_A$), the analytic uncertainty for $A$ obtained with Eq.\,\ref{errA} ($\sigma^{\mathrm{analytic}}_{A}$), the average among the 50 estimated amplitudes ($<A>$), and the uncertainty for the average recovered amplitude ($\sigma_{<A>}$) obtained as in Eq.\,\ref{eq:sigma_a_med}. The atmospheric template has been fitted from simulations containing or not the CMB dipole component, as indicated in the third column of the table.}
\label{tab_fit}
\end{table}

We report here the results of the template function fitting performed by our implementation of the code (Sec.~\ref{sec_fitfunc}). We tested this technique with two templates: a TOD projection of the CMB dipole, and a stable plane-parallel atmosphere of the type $1/\sin(\textbf{el})$ (see Sec.~\ref{sec_template} for a more detailed description of the templates). 

With this aim, we used a set of realistic simulations of intensity at $11$\,GHz, which contain the sky signal, the CMB solar plus orbital dipole, the CMB anisotropies, and N=50 independent realizations of noise, for two cases: white noise only or white plus \oof noise, simulated using the values in Table~\ref{Table_noise_sim}, with the noise generator described in Sec.~\ref{Noise}. These simulations can be directly used for the fitting of the CMB solar plus orbital dipole, while for the fitting of the atmospheric component, we added the term $1$\,K$/\sin(\textbf{el})$, with an amplitude $A_{\text{atmo}}=1$ representing the temperature of the atmosphere at the zenith in units of Kelvin. In addition, in order to test possible degeneracy at the TOD level between the atmospheric and CMB dipole, we used simulations excluding the CMB dipole and including the atmosphere, other than the Galactic signal and the noise. 
By construction, the expected value of the fitted amplitude of the CMB dipole is $A_{\text{d}}=1$, and for the atmosphere it is $A_{\text{atmo}}=1$.  

We report in Table~\ref{tab_fit} the average of the results obtained for the atmospheric and CMB dipole fitting, using the results from the 50 realizations mentioned above. We also report different estimates of the uncertainty:  the uncertainty for a single estimate of $A$ obtained as the dispersion\footnote{Sample variance: $\sigma = \sqrt{\sum_{i=1}^N(A_i-<A_i>)^2/(N-1)}$, being $<\cdot>$ the average of the estimated amplitudes $A$, and $N$ the number of samples.} of the Monte Carlo (MC) realizations ($\sigma^{\rm MC}_A$), the analytic uncertainty for $A$ obtained with Eq.\,\ref{errA} ($\sigma^{\mathrm{analytic}}_{A}$), and the uncertainty for the average recovered amplitude ($\sigma_{<A>}$), which  under the assumption of independent realizations is: 
\begin{equation}
    \sigma_{<A>} = \frac{\sigma^{\text{MC}}_A}{\sqrt{N}}.
    \label{eq:sigma_a_med}
\end{equation}
However, the realizations are not totally independent: although we inject independent noise realizations in the simulations, we always adopt the same data-set, sky signal, geometry of the observations, and  data flagging. Therefore, the simulation are partially correlated, and the final uncertainty $\sigma_{<A>}$ could be slightly underestimated. On the other hand, the fact that the number of simulations is small may induce an overestimated uncertainty as computed from the dispersion. 

Under the assumption that $\sigma_{<A>}$ provides a reliable estimate of the uncertainty of the average recovered amplitudes, we can observe if the method suffers bias effects. In the simulations with only white noise, the atmospheric pattern is perfectly recovered if the CMB dipole is not included in the simulations ($<A_{\rm atmo}>=1.0000\pm0.0001$), while it shows a bias of $\sim 0.38\%$ ($<A_{\rm atmo}>=1.0038\pm0.0001$) if the TODs include also the CMB dipole. This is possibly due to a degeneracy between the atmospheric and the CMB dipole templates. Also the fit of the CMB dipole ($<A_{\rm d}>=0.992\pm0.002$) shows a small bias of $\sim 0.8\%$, which is possibly due to a degeneracy between the Galactic and CMB dipole.  However, in the realistic case with white+$1/f$ noise these biases are totally absorbed by the uncertainties of the fit.
When $1/f$ noise is included in the simulations, the average estimated amplitude $<A>$ is compatible with the expected value $A=1$ within 0.5\,$\sigma$ for the atmosphere fitted from TODs without CMB dipole, within 1.6\,$\sigma$ for the atmosphere fitted when also the CMB dipole is present in the simulations, and within 0.05\,$\sigma$ for the fit of the CMB dipole. 

Finally, we assess the uncertainty of one single estimate of the amplitude $A$, which is the uncertainty to quote when we apply the fitting procedure with the real data (for which only one realization of the data-set is available). We quote the uncertainty on $A$ as the dispersion of the results obtained with different MC realizations, $\sigma^{\text{MC}}$ (fourth column in Tab.\,\ref{tab_fit}; where MC indicates that the uncertainty is obtained as the standard deviation of Monte Carlo realizations). This estimate of the uncertainty can be compared with the analytical one obtained with Eq.\,\ref{errA}, which is reported in the fifth column in  Tab.\,\ref{tab_fit}. We can notice that $\sigma^{\mathrm{analytic}}_{A}$ provides a good estimate of the uncertainty on $A$ in the white noise only case, it being consistent with Monte Carlo estimate $\sigma^{\rm MC}_{A}$, while it underestimates the uncertainty on $A$ when including also $1/f$ noise. Indeed,  the fitting methodology relies on the assumption that the baselines perfectly subtract the $1/f$ component, solving the $\chi^2$ problem (Eq.\,\ref{eq:X_impl}) for a residual TOD dominated by white noise. However, particularly in intensity, we have shown that residual $1/f$ noise is present, and it affects the precision of the amplitude determination. In the case with $1/f$ noise, a more realistic uncertainty is given by the standard deviation of the ten Monte Carlo measurements of $A$, $\sigma^{\text{MC}}_A$, which accounts for the injected noise into the data. 

The result of this analysis, for the realistic case including $1/f$ noise, is that an atmospheric component of the type $1/\sin(\textbf{el})$ can be recovered by our technique with an uncertainty $\sigma^{\text{MC}}_{A_{\text{atmo}}} =$ 0.02, which is a precision of $\sim$ 2\%. On the other hand, for the CMB dipole, we can reach a precision of $\sim$28\%, being $\sigma^{\text{MC}}_{A_{\text{d}}} =0.28$. There are various factors that can explain the difference in the precision achieved for $A_{\text{atmo}}$ and $A_{\text{d}}$, despite the atmosphere and the CMB dipole seems to introduce in the maps the same level of the fluctuations, of the order of few mK (depending on the elevation in the case of the atmosphere). First, the telescope scans one full period  of the atmospheric fluctuation in only one scan. For example, in one single ring-like scan at $el=60^{\circ}$, which takes $30$\,s in the scanning strategy of the wide-survey, the atmospheric pattern is entirely measured, and the amplitude of the fluctuation is of the order of few mK (see Sec.~\ref{sec_atmo}). The CMB dipole, instead, given the QUIJOTE latitude on Earth and the scanning strategy, is scanned from its maximum to its minimum (which are approximately $\pm3$\,mK on a map) with several hours of separation, because, in order to measure it, we have to wait its transit across the sky. As a consequence, the complete measurement of the CMB dipole is spread in time, and the variations introduced in the TOD by the CMB dipole in one single ring scan are much smaller than its peak amplitude of $\approx 3$\,mK, and are therefore also smaller than the atmospheric fluctuations in one scan. In addition, the large \oof noise drifts over a slowly varying template such as that of the CMB dipole complicate the action of recovering its amplitude at the TOD level. Moreover, the CMB solar dipole component is degenerate with the sky map. The orbital CMB dipole allows to break the degeneracy, but its amplitude is approximately 10 times smaller than that of the solar component. This consideration provides one more explanation of why the level of precision of the CMB dipole fitting is lower than that of atmospheric template, with this technique. The CMB dipole can be better fitted directly from the map. 

\subsection{Cross-correlations with the CMB} \label{sec_cmb}

\begin{figure}
    \centering
    \includegraphics[width=0.49\textwidth]{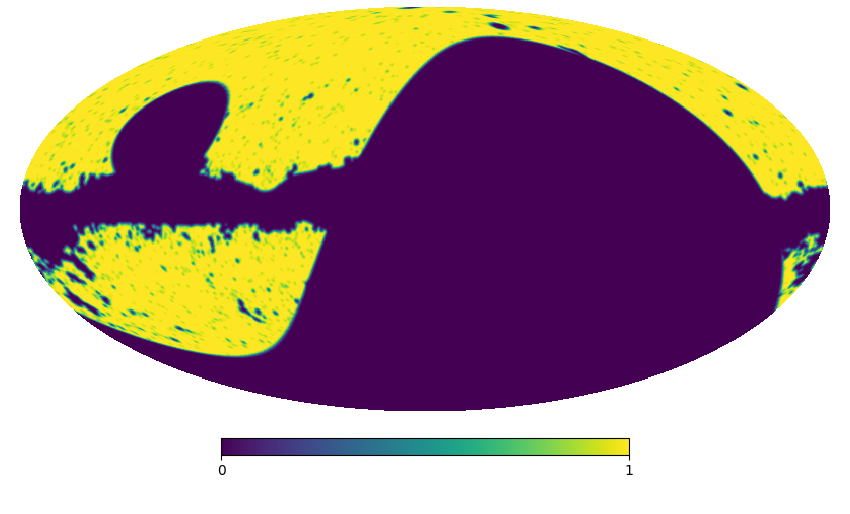}
    \caption{Mask adopted for the cross-correlations with the CMB. The mask is obtained from the Planck confidence CMB mask for temperature \citep{pla2018comp_sep} apodized with a simple 2-degrees smoothing.
    }
    \label{CMB_mask}
\end{figure}

Although the primary CMB anisotropies are not the dominant signal in the QUIJOTE-MFI wide-survey intensity maps (see Fig.~\ref{fig_cl}) they could be detected via cross-correlations. The simulated maps that are presented in this work can be used to test the stability of the CMB detection method that is applied to the real data, and to support the result that will be presented in Rubi{\~n}o-Mart{\'{\i}}n et al. (in prep).

The method is based on the computation of the level of correlation of the QUIJOTE maps $\textbf{m}_{\rm QJT}$ with the CMB anisotropies maps $\textbf{m}_{\rm CMB}$ as traced by Planck data, accounting simultaneously for the chance alignment between the CMB and the Galactic foregrounds. We assume that the QUIJOTE map of a given horn and frequency  is a linear combination of the CMB map, of a template of Galactic foregrounds \textbf{f}, and of the noise \textbf{n}, as:
\begin{equation}
    \textbf{m}_{\rm QJT} = \alpha \cdot \textbf{m}_{\rm CMB} + \beta \cdot \textbf{f} + \textbf{n}
\end{equation}
where $\alpha$ and $\beta$ are the parameters of the linear combination of the CMB and the foregrounds map, respectively. Let us perform a cross-correlations of the QUIJOTE map with the CMB and with the foregrounds map, which gives:
\begin{equation}
\begin{cases}
    C_\ell^{{\rm QJT \times CMB}} = \alpha \cdot C_\ell^{{\rm CMB \times CMB}} +\beta \cdot C_\ell^{\rm f \times CMB}\\
      C_\ell^{\rm QJT \times f} = \alpha \cdot C_\ell^{\rm CMB \times f} +\beta \cdot C_\ell^{\rm f \times f}
\end{cases}
\label{eq_cmb_system}
\end{equation}
where $C_\ell^{X \times Y}$ is the cross power spectrum of map $X$ and map $Y$. In Eq.~\ref{eq_cmb_system}, we assumed that the noise map of QUIJOTE does not play any role in the cross-correlations, and that the parameters $\alpha$ and $\beta$ do not change with the angular scale. By solving this system of equations with respect to $\alpha$ we get:
\begin{equation}
    \alpha =  \left <\frac{C_\ell^{{\rm QJT \times CMB}}}{C_\ell^{{\rm CMB \times CMB}}}-\frac{C_\ell^{{\rm QJT \times f}}}{C_\ell^{{\rm f \times f}}}\cdot \frac{C_\ell^{{\rm CMB \times f}}}{C_\ell^{{\rm CMB \times CMB}}}\right >_{\ell\in[100,200]}
    \label{eq_A_CMB}
\end{equation}
where the brackets $<\cdot>$ represent an average within all multipoles in the range $\ell\in[100,200]$, so in proximity of the first peak of the CMB angular power spectrum. 
This range of multipoles is a particular selection in which, first,  the CMB power spectrum is in the signal dominated regime with respect to the signal error of the QUIJOTE maps (as shown in Fig.~\ref{fig_cl}), and second,  any residual from the dipole mode coupling which is left by the power spectrum estimator is negligible  (see discussion in Sec.~\ref{sec_cl}).  If the CMB anisotropies are correctly recovered, and the QUIJOTE maps are properly calibrated, we expect to measure with Eq.~\ref{eq_A_CMB} a value of $\alpha=1$, which can be read as the amplitude of the CMB anisotropies map measured by QUIJOTE.

\begin{figure}
    \centering
    \includegraphics[width=0.49\textwidth]{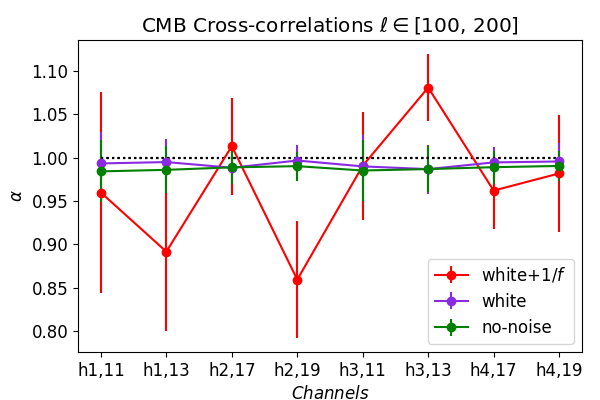}
    \caption{Amplitude $\alpha$ of the CMB in the simulated QUIJOTE-MFI maps, obtained with cross-correlations. The analysis is performed with simulations containing white plus \oof noise (red), white noise (purple) and no-noise (green). The error bars for the measurement of $\alpha$ are obtained with rotations of the CMB map.}
    \label{fig:CMB_crosscorr}
\end{figure}

\begin{table}
\centering
\begin{tabular}{cccc}
\hline\hline
Channel &  $\alpha$ (no-noise)  &  $\alpha$ (white noise) & $\alpha$ (white+\oof{}noise) \\ \hline\hline
h1,11 & 0.98 $\pm$ 0.04 & 0.99 $\pm$ 0.04 & 0.96 $\pm$ 0.12 \\
h1,13 & 0.99 $\pm$ 0.03 & 0.99 $\pm$ 0.03 & 0.89 $\pm$ 0.09 \\ \hline
h2,17 & 0.99 $\pm$ 0.02 & 0.99 $\pm$ 0.02 & 1.01 $\pm$ 0.06 \\
h2,19 & 0.99 $\pm$ 0.02 & 1.00 $\pm$ 0.02 & 0.86 $\pm$ 0.07 \\\hline
h3,11 & 0.99 $\pm$ 0.04 & 0.99 $\pm$ 0.04 & 0.99 $\pm$ 0.06 \\
h3,13 & 0.99 $\pm$ 0.03 & 0.99 $\pm$ 0.03 & 1.08 $\pm$ 0.04 \\\hline
h4,17 & 0.99 $\pm$ 0.02 & 0.99 $\pm$ 0.02 & 0.96 $\pm$ 0.04 \\
h4,19 & 0.99 $\pm$ 0.02 & 1.00 $\pm$ 0.02 & 0.98 $\pm$ 0.07 \\\hline\hline
\end{tabular}
        \caption{Amplitude $\alpha$ of the CMB in the simulated QUIJOTE-MFI maps, obtained with cross-correlations. The analysis is performed with simulations without noise (first column), with white noise only (second column), and with white plus \oof noise (third column). The error bars are obtained with rotations of the CMB map.}
        \label{tab:cmb_crosscorr}
\end{table}

We performed this analysis  with the simulated intensity maps of the QUIJOTE-MFI wide-survey, smoothed to 1-degree, and for the case with no-noise, white noise, and white plus \oof noise. It can be noticed from Fig.~\ref{fig_cl} that, while in the cases of no-noise and white noise the CMB can be detected with high signal to noise, the power of the residual \oof noise is above the level of the CMB at all multipoles. Consequently, the detection of the CMB with \oof is more affected by the noise, as expected. 

With the real data, a reasonable foreground template that can be used for this analysis is the K-band map of WMAP (\citealt{WMAPmaps}), after subtracting the CMB component. Therefore, here we use as a foreground template the simulated foreground map at 11\,GHz, scaled to 22.8\,GHz with a temperature spectral index of $-3$.

The cross power spectra are computed with \xpol, using the Planck confidence mask for temperature \citep{pla2018comp_sep}, which excludes the low confidence regions of the CMB map from the sky observed by QUIJOTE. The mask is apodized with a simple 2 degrees smoothing, and it is shown in Fig.~\ref{CMB_mask}.

We estimate the uncertainty of the parameter $\alpha$ with rotations of the CMB maps. We perform 19 rotations of the CMB in galactic longitude, with $\Delta l =18$\,deg, and we estimate with Eq.~\ref{eq_A_CMB}, for each rotation angle, the rotated amplitude $\text{rot}(\alpha)$. The expected correlation of the QUIJOTE map with a rotated CMB map is zero, so $<\text{rot}(\alpha)>=0$, and the standard deviation of the distribution can be used as an estimate of the uncertainty on $\alpha$.
 

The results are reported in Tab.~\ref{tab:cmb_crosscorr} and are shown in Fig.~\ref{fig:CMB_crosscorr}, where we represent the amplitude $\alpha$ of the correlation between all channels of the simulated QUIJOTE-MFI maps and the CMB. The analysis is performed with no-noise (green), white noise (purple) and white plus \oof noise simulations (red). We notice that, in the cases of no-noise and white noise only, we detect the  CMB with a precision of $\approx  2-4\%$. When we include the \oof noise, we recover the CMB with a precision of $\approx 4-12\%$, meaning that, despite the noise, we obtain $\approx 10-20\,\sigma$ detection of the CMB, depending on the QUIJOTE channel.  We obtain, as expected, zero correlation with the rotated CMB map.

\section{Conclusions} \label{sec_conclusions}
We presented \namecode{}, a map-making code implemented for the construction of the maps of the MFI instrument of the QUIJOTE experiment. \namecode{} is based on the destriping algorithm with priors on the baselines for the suppression of the \oof{}noise, and implements a technique to fit for a general template at the map-making level. This feature is particularly useful for the analysis of ground-based CMB experiments. 

We performed simulations of the QUIJOTE-MFI wide-survey (Rubi{\~n}o-Mart{\'{\i}}n et al. in prep) to test the performance of \namecode{}. We showed a realistic simulated version of the QUIJOTE-MFI intensity and polarization wide-survey maps (Fig.~\ref{fig_maps}), at 11, 13, 17 and 19\,GHz. We then conducted a detailed analysis of the simulated maps at the map level and at the power spectrum level, with special emphasis on the stability of the reconstruction of the large angular scales. \namecode{} is able to reconstruct the CMB dipole with $\sim 0.1\,\%$ accuracy, at the map level.

We presented the angular power spectra of the simulated maps with no-noise, white noise, and realistic white plus \oof noise (Fig.~\ref{fig_cl}). We studied the signal error and the transfer function of the map-making code, in combination with the scanning strategy of the wide-survey
. \namecode{} performs well at all angular scales: the signal error is lower than 0.001\% at multipoles in $20<\ell<200$, for TT, EE and BB (Fig.~\ref{fig_signal_error}), at all the QUIJOTE-MFI frequencies. 
Furthermore, the results obtained for the transfer function (Fig.~\ref{fig_transfer_function_nonoise}) show that \namecode{} performs a perfect reconstruction of the sky signal at multipoles  $\ell>10$ in TT and $\ell>8$ in EE and BB, for the partial sky coverage of the MFI wide survey. Moreover, in polarization, also the larger angular scales  $2<\ell<8$ are precisely recovered, within 2\,\% error.  

Afterwards, we tested the template fitting procedure that is implemented in \namecode{}, using a template of the atmosphere and of the solar plus orbital CMB dipole (Sec.~\ref{sec_fit_results}). For the noise levels in the QUIJOTE-MFI wide survey, we expect to recover the amplitude of the atmospheric fluctuations with a precision of $2\,\%$, and of 28\,\% for the CMB dipole. 

Finally, we presented a validation with simulations of the technique that is applied for the detection of the CMB intensity anisotropies from the real wide-survey data. Even with \oof{}noise, the QUIJOTE-MFI wide-survey could detect the CMB at $10-20\,\sigma$. 

Although \namecode\ is specifically implemented for the QUIJOTE MFI instrument, it is based on totally general principles. The code is being adapted to be used with other QUIJOTE instruments. We encourage the reader interested to use this code to contact the authors.


\section*{Acknowledgements}

The QUIJOTE experiment is being developed by the Instituto de Astrofisica de Canarias (IAC), the Instituto de Fisica de Cantabria (IFCA), and the Universities of Cantabria, Manchester and Cambridge. Partial financial support is provided by the Spanish Ministry of Science, Innovation and Universities
under the projects AYA2007-68058-C03-01, AYA2010-21766-C03-02, AYA2014-60438-P, AYA2017-84185-P, IACA13-3E-2336, IACA15-BE-3707, EQC2018-004918-P, the Severo Ochoa Program SEV-2015-0548, and also by the Consolider-Ingenio project CSD2010-00064 (EPI: Exploring the Physics of Inflation).
This project has received funding from the European Union's Horizon 2020 research and innovation program under grant agreement number 687312 (RADIOFOREGROUNDS).
This research used resources of the National Energy Research Scientific Computing Center, which is supported by the Office of Science of the U.S. Department of Energy under Contract No. DE-AC02-05CH11231.
This research made use of computing time available on the high-performance computing systems at the IAC. We thankfully acknowledge the technical expertise and assistance provided by the Spanish Supercomputing Network (Red Espa\~nola de Supercomputaci\'on), as well as the computer resources used: the Deimos/Diva Supercomputer, located at the IAC. 
RBB and JDBA acknowledge the Spanish Agencia Estatal de Investigaci{\'o}n (AEI, MICIU) for the financial support provided under the projects with references PID2019-110610RB-C21, ESP2017-83921-C2-1-R and AYA2017-90675-REDC, co-funded with EU FEDER funds, and also acknowledge the funding from Unidad de Excelencia Mar{\'i}a de Maeztu (MDM-2017-0765). RBB and JDBA also acknowledge the Santander Supercomputaci{\'o}n support group at the Universidad de Cantabria who provided access to the Altamira supercomputer at the Instituto de F\'{\i}sica de Cantabria (IFCA-CSIC), member of the Spanish Supercomputing Network for running the QML code.

\section*{Data Availability}
The data and the code presented in this paper are not publicly available, but they can be shared on reasonable request to the corresponding authors.


\bibliographystyle{mnras}
%
%
\bibliography{biblio,quijote} 


\bsp	
\label{lastpage}
\end{document}